\documentstyle[onecolumn]{mn}    
\def\lsim{~\rlap{$<$}{\lower 1.0ex\hbox{$\sim$}}}
\def\bsim{~\rlap{$>$}{\lower 1.0ex\hbox{$\sim$}}}

\def\kms{\ {\rm Km\,s^{-1}}}
\def\hmpc{\ {\rm h^{-1}Mpc}}

\def\dd{{\rm d}}
\def\ln{{\rm ln}}
\def\pa{\partial}

\def\pmb#1{\setbox0=\hbox{#1}%
\kern-.025em\copy0\kern-\wd0
\kern.05em\copy0\kern-\wd0
\kern-.025em\raise.0433em\box0}

\def\vv{\pmb{$v$}}
\def\valpha{\pmb{$\theta$}}

\def\vtheta{\pmb{$\theta$}}

\def\vg{\pmb{$g$}}

\def\vx{\pmb{$x$}}
\def\ve{\pmb{$e$}}
\def\vr{\pmb{$r$}}
\def\vs{{\pmb{$s$}}}

\def\vC{\pmb{$C$}}

\def\da {{\dot a}}

\def\qn{{q_{_n}}}
\def\pn{{p_{_n}}}
\def\etal{{\it et al.\ }}

\def\cpar{{\vC^\parallel}}

\begin{document}
\title[Least Action  in Redshift Space]
{Peculiar  Velocity Reconstruction with Fast Action Method: \\
Tests on Mock Redshift Surveys}                                          
\author[Branchini, Eldar \& Nusser]{Enzo Branchini$^1$, Amiram Eldar$^{2,3}$ \& Adi Nusser$^3$ 
\\
$^1$ Dipartimento di Fisica,
Universit\'a di Roma TRE,
Via della Vasca Navale 84, 00146, Roma, Italy \\
$^2$ Racah Institute of Physics, The  Hebrew University,
Jerusalem 91904, Israel \\   
$^3$ Physics Department,
Technion, Haifa 32000, Israel \\
}
\maketitle

\begin{abstract}
  
  We present extensive tests of the Fast Action Method (FAM) for
  recovering the past orbits of mass tracers in an expanding universe
  from their redshift-space coordinates at the present epoch.  The
  tests focus on the reconstruction of present-day peculiar velocities
  using mock catalogs extracted from high resolution $N$-body
  simulations.  The method allows for a self-consistent treatment of
  redshift-space distortions by direct minimization of a modified
  action for a cosmological gravitating system.  When applied to
  ideal, volume limited catalogs, FAM recovers unbiased peculiar velocities
  with a 1-D,  $1\sigma$ error of $\sim 220 \kms$, if velocities are smoothed on a scale of 
  $5 \hmpc$. Alternatively, when no smoothing is applied, FAM predicts nearly unbiased 
  velocities for objects residing outside the highest density regions.
  In this second case the $1\sigma$ error decreases to 
  a level of $\sim 150 \kms$. The correlation
  properties of the peculiar velocity fields are also correctly
  recovered on scales larger than $5 \hmpc$.  Similar results are
  obtained when FAM is applied to flux limited catalogs mimicking the
  IRAS PSC$z$ survey.  In this case FAM reconstructs peculiar
  velocities with similar intrinsic random errors,  while
  velocity-velocity correlation properties are well reproduced beyond
  scales of $\sim 8\hmpc $. We also show that  
  FAM provides better velocity 
  predictions than other, competing methods based on linear theory or 
  Zel'dovich approximation.
  These results indicate that FAM can be
  successfully applied to presently available galaxy redshift surveys
  such as IRAS PSC$z$.

\end{abstract}


\section{Introduction}
\label{sec:intro}       

In the standard paradigm, the formation of cosmological structures is
driven by gravitational amplification of tiny initial density
fluctuations (e.g. Peebles 1980).  In addition to gravity,
hydrodynamical processes can greatly influence the formation and
evolution of galaxies, groups and clusters of galaxies.
Hydrodynamical effects, however, play a minor role in shaping the
observed distribution of galaxies on scales a few times larger than
the size of galaxy clusters. Therefore, gravitational instability
theory directly relates the present-day large scale structure to the
initial density field and provides the frame-work within which the
observations are analyzed and interpreted.  Gravitational instability
is a non-linear process.  Analytic solutions exist only for
configurations with special symmetry, and approximate tools are
limited to moderate density contrasts.  So, numerical methods are necessary
for a full understanding of the observed large scale
structure of the universe.  There are two complementary numerical approaches.  The
first approach relies on $N$-body techniques designed to solve an
initial value problem in which the evolution of a self-gravitating
system of massive particles is determined by numerical integration of
the Newtonian differential equations.  Combined with semi-analytic
models of galaxy formation,  $N$-body simulations have become an
essential tool for comparing the predictions of cosmological
models with the observed properties of galaxies.  Because the exact
initial conditions are unknown, comparisons between simulations and
observations are mainly concerned with general statistical properties.
The second approach aims at finding the past orbits of mass tracers
(galaxies) from their observed present-day distribution, independent
as much as possible of the nature of the dark matter.  The orbits must
be such that the initial spatial distribution is homogeneous.  This approach
can be very useful for direct comparisons between different types of
observations of the large scale structure.  Most common are the
velocity-velocity (hereafter {\it v-v}) comparisons between the
observed peculiar velocities of galaxies are and the velocity field
inferred from the galaxy distribution in redshift surveys (Davis,
Nusser \& Willick 1996, Willick \etal 1997, Willick \& Strauss 1998,
Nusser \etal 2000, Branchini \etal 2001) Alternatively, one can also
perform a density-density  comparison between the
mass density field inferred from the peculiar velocities of galaxies
via POTENT-like methods (Bertschinger \&
Dekel 1989) or Wiener Filtering techniques (Zaroubi \etal 1995)
and  the observed galaxy distribution (see Sigad \etal 1998 
and references therein).  Both
types of analysis yield the cosmological mass density parameter
$\Omega_m$, for an assumed biasing relation between the distribution
of galaxies and the mass density field. 
Any systematic mismatch between the fields 
serves as an indication to the nature of
galaxy formation and/or the origin of galaxy intrinsic scaling
relations used to measure the distances, provided that errors
in the calibration have been properly corrected for.  
This second approach also allows to
perform back-in-time reconstructions of the density field on scales
$\sim 5 \hmpc$ both in real (e.g, Nusser \& Dekel 1992, Gramman 1993,
Croft \& Gazta\~{n}aga 
1998, Frisch \etal 2001) 
and redshift space (Narayanan \& Weinberg
1998, Monaco \& Efstathiou 1999).  Some of these methods
has been applied to recover the past orbits from all-sky
galaxy redshift surveys (Monaco \etal 2000, Narayanan \etal 2001).

Finding the orbits that satisfy initial homogeneity and match the
present-day distribution of mass tracers is a boundary value
problem. This problem naturally lends itself to an application of
Hamilton's variational principle where the orbits of the objects are
found by searching for stationary variations of the action subject to
the boundary conditions.  The use of the Principle of Least Action in
a cosmological frame-work has been pioneered by Peebles (1989) and 
has long been restricted to small systems such as the Local Group
(Peebles 1990, Peebles 1994, Schmoldt \& Saha 1998, Sharpe \etal 2001)
and the Local Supercluster (Shaya, Peebles \& Tully 1995). Early
applications to large galaxy redshift surveys have been hampered by
the computational cost of handling the relatively large number of
objects.  Subsequent numerical applications speeded
up the method and allowed the reconstruction of the orbits of $\sim
10^3$ particles (Shaya, Peebles \& Tully 1995).  However, it was only
recently that the improvement of the minimization techniques and the
use of efficient gravity solvers made it possible to deal with more
than $10^4$ objects (e.g. the Fast Action Method [hereafter FAM] by
Nusser \& Branchini 2000 and the Perturbative Least Action Method by
Goldberg \& Spergel 2000), comparable to the number of objects
contained in the largest all-sky galaxy redshift catalogs presently
available such as the Optical Redshift Survey (Santiago \etal 1995,
1996) and IRAS PSC$z$ (hereafter  PSC$z$, Saunders 1996, Saunders \etal 2000).  The
situation has also been improved thanks to the invention of
self-consistent direct schemes for accounting for redshift
distortions, arising from the systematic differences between the
distribution of galaxies in real and redshift space (e.g. Kaiser
1987).  Previous applications (e.g., Shaya, Peebles \& Tully 1995)
relied on iterations. However, self-consistent treatments of the
problem have been proposed by Schmoldt \& Saha (1998), Nusser \&
Branchini (2000), Phleps (2000), Susperregi (2001) and Goldberg
(2001).

In this work we will implement and test one of the Numerical Action
Methods on realistic large mock redshift catalogs to reconstruct
galaxy orbits over a large region of the universe.  More precisely we
will extend the Fast Action Method of Nusser \& Branchini (2000, hereafter
NB) to redshift space and focus on the problem of predicting galaxy
peculiar velocities from a flux-limited, all-sky redshift catalog
resembling the PSC$z$ one.

The outline of the paper is as follows.  In Section~\ref{sec:method}
we review FAM in real and redshift space The performance of FAM is
evaluated in the case of the ideal spherical infall in
Section~\ref{sec:sphere}.  Then a suite of more demanding tests is
performed using the mock catalogs of galaxies described in
Section~\ref{sec:mock}.
FAM is then applied to ideal, volume-limited mock catalogs in
Section~\ref{sec:Nbodyvol} and to flux-limited mock catalogs in
Section~\ref{sec:Nbodyflux}.  Particular emphasis is given to the
ability of FAM in returning an unbiased estimate of galaxy velocities
and their correlation properties. The main conclusions are discussed
and summarized in Section \ref{sec:disconc}.  We will use the terms
FAMz and FAMr to refer to implementations of FAM in redshift and real
distance space, respectively. Statements referring to FAM are meant to
apply to both FAMz and FAMr. For brevity we will use x-space and
s-space to refer to real-space and redshift-space, respectively.

\section{The Fast Action Method}
\label{sec:method}       

In this Section we briefly summarize the Fast Action Method.  For a detailed
description of the method we refer the reader to Sections 2 and 4.2 of
NB.  We follow the standard notation in which $a(t)$ is the scale
factor, $H(t)=\da /a$ is the Hubble function, $\Omega_m=\bar
\rho/\rho_c$ is the ratio of the background matter density of the universe, $\bar
\rho$, to the critical density, $\rho_c=3H^2/8\pi G$.
We denote the comoving coordinate of a patch of matter by $\vx $, and
the corresponding comoving velocity by $\vv=\dd \vx /\dd t$.  Also,
let $D(t)$ be the linear density growing mode normalized to unity at the
present epoch, and $f(\Omega_m)=\dd \ln D/\dd \ln a \approx
\Omega_m^{0.6}$ (e.g., Peebles 1980).  In the following we will use
$D(t)$ as the time variable in the equations of motion, with $\vtheta=\dd
\vx/\dd D$ as the corresponding velocity.  Expressed in term of the
time variable $D$, the equations of motion are almost independent on $\Omega_m$
and the cosmological constant (Gramann 1993, Mancinelli \& Yahil
1995, Nusser \& Colberg 1998).

The evolution of a cosmological self-gravitating, isolated system of
$N$ equal mass particles in a volume $V$ is governed by the following
equations of motion,
\begin{equation}
\frac{\dd \valpha_i}{\dd D} +\frac{3}{2}\frac{1}{D}\vtheta_i =
\frac{3}{2}\frac{1}{D^2}\frac{\Omega_m}{f^2(\Omega_m)}\vg(\vx_i) , 
\label{eq:euler}
\end{equation}
where the subscript $i=1\cdots N$ is the particle index, and  $\vg$
is the peculiar gravitational force field per unit mass.
If the  particles are unbiased tracers of the underlying density
field then,
\begin{equation}
\vg(\vx)=-
\frac{1}{4\pi  \bar n }\sum_i \frac{\vx-\vx_i} {|\vx-\vx_i|^3} 
+\frac{1}{3}\vx \, , 
\label{eq:pot}
\end{equation}
where $\bar n=N/V$ is the mean number density of particles inside $V$.
In this expression for $\vg$ we have assumed that $N M=\bar \rho V$
where $M$ is the mass of a particle.  These equations of motion can be 
derived from  the action,
\begin{equation}
{\mathrm S}=\int_0^{1} \dd D   \sum_i
\left\{\frac{1}{2}  D^{3/2} \vtheta_i^2
+\frac{3}{2}\frac{1}{D^{1/2}}\frac{\Omega_m}{f^2(\Omega_m)} 
\left[  \frac{1}{4\pi \bar n}
\sum_{j<i}\frac{1}{|\vx_i-\vx_j|}
+\frac{ \vx_i^2}{6}  \right]   \right\} \, 
\label{eq:lapd}
\end{equation}
under stationary first variations of the orbits that leave $\vx $
fixed at the present epoch
and satisfy the constraint $D^{3/2}\vtheta\rightarrow 0$ as
  $D\rightarrow 0$ (Peebles 1989, NB). The second condition on the
  velocities guarantees homogeneity as $D \rightarrow 0$. FAMr
  expands the orbits in a time dependent base functions $\qn(D)$ in
  the form,
\begin{equation}
\vx_i(D)=\vx_{i,0}+\sum_{n=1}^{n_{max}} \qn(D) \vC_{i,n}  ,
\label{eq:expand} 
\end{equation}
where $\vx_{i,0}$ is the position of the particle $i$ at the present epoch, and
the vectors $\vC_{i,n}$ are the expansion coefficients with respect to
which the action is varied, i.e., they satisfy $\pa {\mathrm S}/\pa
\vC_{i,n}=0$.  The base functions $\qn$ and their derivatives
$\pn(D) \equiv \dd \qn/\dd D$ are linear combinations of $(1-D),
(1-D)^2\cdots (1-D)^{n_{max}}$ so that the particle positions at $D=1$
is fixed, and $\lim_{D\rightarrow 0} D^{3/2} q_n(D) \vtheta(D)=0$
ensures initial homogeneity.  To simplify the expression of the
gradient of the action, $\pa {\mathrm S}/\pa \vC_{i,n}$, NB imposed
following the orthonormality condition on the functions
$\pn(D)$,
\begin{equation}
\int_0^1 \dd D D^{3/2} \pn(D) p_m(D)=\delta^{\rm K}_{m,n}, 
\label{eq:ortho}
\end{equation}
where $\delta^{\rm K}$ is the Kronecker delta function.  
The expansion of the orbits given in (\ref{eq:expand}) has to be modified 
when the constraints are galaxy redshifts, rather than
the true distances.  
The redshift coordinate  of an object  is
\begin{equation}
\vs_{_0}=H_0 \vx_{_0}+ \left(\vv_{_0}\cdot \hat \vs_{_0} \right) \hat \vs_{_0}, 
\label{eq:red}
\end{equation}
where the subscript 0 refers to quantities at the present time, and
$\hat \vs_0$ is a unit vector directed along the line of sight to
the galaxy.  When the constraints are the particle positions
in s-space, $\vs_{_{i,0}}$, the appropriate expansion of the orbits
can be written as,
\begin{equation}
\vx_{i}(D)=H_0^{-1}\vs_{{_{i,0}}} +
\sum_n \qn(D) \vC_{i,n} -f_0 \left( \sum_n p_{n,0} \cpar_{i,n} \right)  
{\hat \vs}_{_{i,0}},
\label{eq:zorbit}
\end{equation}
where  $p_{n,0}=\pn(D=1)$, and the symbol $(^\parallel)$
indicates the component of a vector in the direction of line
of sight at the present epoch, $\hat \vs_{_{i,0}}$.  The trivial
dependence on $H_0$ can be completely eliminated by working with $H_0
\vx$ instead of $\vx$.  With this expression for the orbits, the
gradient of the action subject to the new boundary conditions is

\begin{equation}
\pa {\mathrm S}/\pa \vC_{i,n}={\rm {\bf I}}_{i,n} -f_0 p_{n,0}\left(
\valpha_{{_{i,0}}}^\parallel+{\rm {\bf I}}^\parallel_{i,n} \right),
\label{eq:zgrad}
\end{equation}
where $f_0=f(\Omega_{m,0})$,
\begin{equation}
{\rm {\bf I}}_{i,n} = \int_0^1 \dd D D^{3/2} {q}_n 
\left[\frac{\dd \vtheta_i}{\dd D} +
\frac{3}{2}\frac{\vtheta_i}{D} -
\frac{3}{2}\frac{1}{D^2}\frac{\Omega_m}{f^2}\vg_i \right]  
\end{equation}
and
\begin{equation}
{\rm {\bf I}}^\parallel_{i,n} = \int_0^1 \dd D D^{3/2} {q}_n 
\left[\frac{\dd \vtheta^\parallel_i}{\dd D} +
\frac{3}{2}\frac{\vtheta^\parallel_i}{D} -
\frac{3}{2}\frac{1}{D^2}\frac{\Omega_m}{f^2}\vg^\parallel_i \right] .
\label{eq:I1}
\end{equation}

The equations (\ref{eq:zgrad}) 
 differ from the equations of motion
 by the  boundary term, $f_0
p_{{_{n,0}}} \left( \valpha_{{_{i,0}}}^\parallel+ {\rm {\bf I}}^\parallel_{i,n} \right)$.
This term can be  eliminated by adding a kinetic energy term to the action,
as follows
\begin{equation}
{\cal S} = {\mathrm S}  +
\frac{1}{2} f_0 \sum_i \left(\valpha_{{_{i,0}}}^\parallel\right)^2 \, .
\label{lapDN}
\end{equation}
Given the redshifts of the particles, the minimization of the action ${\cal S} $ 
leads to the equations of motions
(eq.~\ref{eq:euler}).  Similar transformations of this type have been
proposed by Schmoldt \& Saha (1998) and Withing (2001).

Our strategy is to find orbits that are as close
as possible to the Hubble flow.  Therefore,  we search for the
minimum of the action and do not look for stationary points which might describe
oscillatory behavior of the orbits (Peebles 1990, 1994).  To find the
coefficients $\vC_{i,n}$ that
minimize the action, FAM uses the Conjugate Gradient Method (CGM)
which is fast and easy to implement (Press \etal 1992).  
The gravitational force $\vg$ and its potential are computed using
the TREECODE gravity solver (Bouchet \& Hernquist 1988). The time
integration in the expression for the action is done using the
Gaussian quadrature method with  10 points at the time abscissa
(Press \etal 1992). 
The CGM requires an initial guess for $ \vC_{i,n} $.  In the standard
FAM application we compute the initial guess using the 
linear theory relation between the velocity and mass distribution.
The minimum of the action proved to be rather insensitive to the
choice of initial guess for $ \vC_{i,n} $, as we have checked by
running  FAM experiments with initial $\vC_{i,n}$ both set
 to zero and  to random numbers with appropriate variance. 
Besides the initial set of $
\vC_{i,n} $, the other free parameters  are the softening used
by the gravity solver and the tolerance parameter that sets the
convergence of the CGM method.

\section{The Spherical Infall Model Test}
\label{sec:sphere}

The collapse of a spherical over-dense perturbation can be followed
analytically into the nonlinear regime before the occurrence of shell
crossing (e.g., Peebles 1980).  Testing FAM with the spherical
collapse model will allow an assessment of the ``shot-noise'' errors
in the recovered velocities. These errors arise from the discrete
sampling of the density field and  can be quantified 
by applying FAM in two different ways.  In the first,
we treat each particle as a uniform spherical shell. So the
gravitational field on a particle is radial and only due to particles
lying within its radius.  In this way particles move along radial
orbits and  shot-noise errors are minimized. 
In the second, each
particle is treated as point mass and the gravitational field is
computed from the TREECODE as in any other application of FAM.  The
comparison between the peculiar velocities in these two different ways
of applying FAM is an important test for assessing the performance of
FAM under general conditions.

Consider a spherical density fluctuation described
by the following radial density profile 
at the initial redshift $ z_{in} =50$: 
\begin{equation}
\delta_{in}(r)  =\delta_0 \left[ 1+\left({{r}\over{r_c}}\right)^2\right]^{-3},
\label{eq:sph}
\end{equation}
where $r_c = 40(1+z_{in})^{-1}\ {\rm Mpc}$,
$\delta_0=(1+z_{in})^{-1}$. The corresponding initial velocity is
obtained using the velocity-density relation of the linear
growing mode. We assume a flat universe with
$\Omega_m=1$ and $\Omega_{\Lambda}=0$. The analytic solution returns the
density and peculiar velocity fields at a generic time. 
The evolved density field is then randomly sampled
with 20,000 particles within $80$ Mpc.  Each particle is then assigned
a peculiar velocity according to the analytic solution and its
redshift coordinate is computed accordingly.  The distribution of
these particles in x-space and s-space are then fed into FAMr and FAMz,
respectively.  We have used $n_{max}=6$ base functions to expand the
orbit of each particle, a softening of $\epsilon=0.5$ Mpc, and a tolerance
parameter $tol =10^{-8}$ to determine the convergence of the CGM
solution.

\begin{figure}
\vspace{13truecm}
{\includegraphics{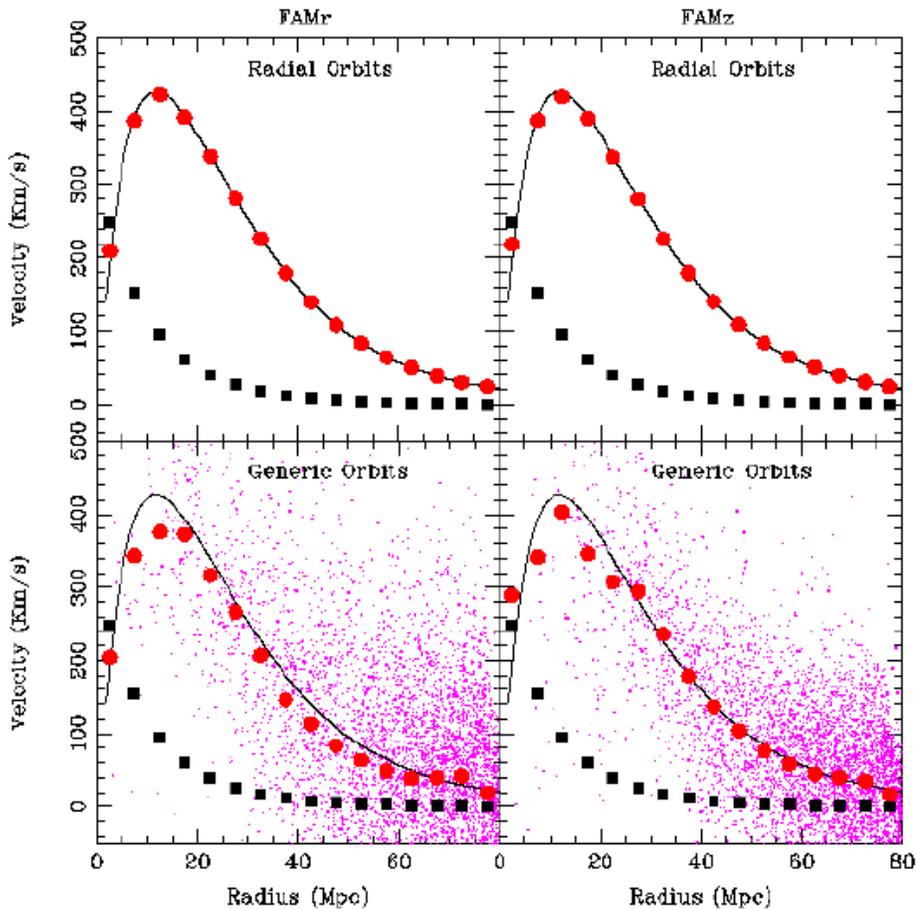}}
\caption{Radial velocity profiles from FAMr (left) and FAMz (right) 
experiments. In all panels the continuous line shows the analytic solution for 
the spherical infall case. Filled dots represent the average FAM
radial velocities measured  in radial shells of 5 Mpc.
The filled squares show the final 
overdensity  profile, $\delta_{fin}(r)\times 100$.
The numerical solution shown in the top panels is obtained 
by imposing radial FAM orbits.
This constraint is removed in the
solutions shown in the two bottom plots, where
the scattered points represent
FAM radial velocities measured at the position of a subset of particles.
}
\label{fig:sph1}
\end{figure}

In the first FAM experiments, particles  move in
radial orbits under the action of the radial gravity force. 
The final radial velocity profiles in the FAMr and FAMz
experiments are shown, respectively, in the top-left and top-right
panels of fig.~\ref{fig:sph1}.  In both panels, the velocity profile in the analytic
solution is represented by the continuous solid line.
The filled dots show the average peculiar velocity computed within
radial shells of 5 Mpc. The scatter around the mean is negligible.
The filled squares represent the final overdensity profile, 
$\delta_{fin}(r) \times 100$, 
also in radial bins of 5 Mpc.
Both FAMr and FAMz  match the analytic 
solution very well at all radii.

We have repeated the FAMz and FAMr experiments with 
the gravity force field  computed by summing
over the 3-dimensional discrete distribution of the point masses.
This generates random shot-noise errors in the FAM recovered velocities.
The radial velocity profiles are shown in the two bottom panels of
fig.~\ref{fig:sph1}.  The results of the FAMr and FAMz experiments are very similar.
The average FAM solutions (filled dots) are
still close to the true one.  FAM appears to underestimate the true
velocities in the high density regions. However, the effect is of
little significance ($\sim 30 \kms$) when compared to the 1-D scatter
around the mean ($\sim 120 \kms$) which, instead, does not seem to
depend on the local density. 

\begin{figure}
\vspace{13truecm}
{\includegraphics{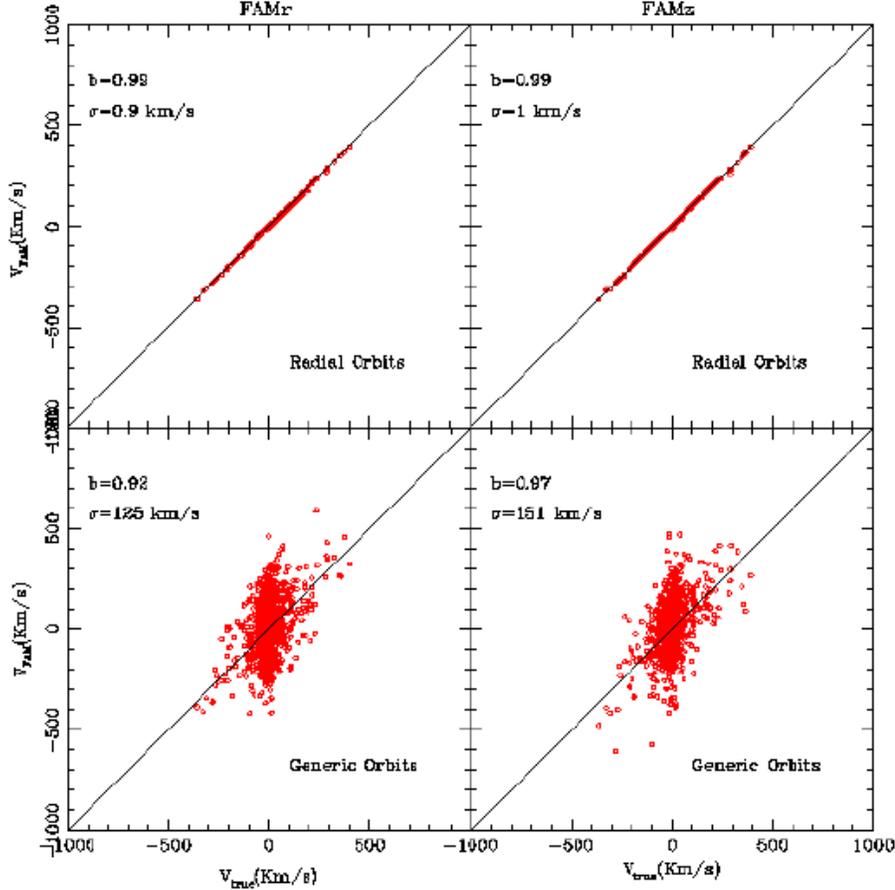}}
\caption{ Cartesian X-components 
of FAMr (left) and FAMz (right) 
velocities vs. true ones. 
Only velocities of $\sim 1000$ randomly selected particles are shown.
The upper panels show the case of purely radial orbits.
The lower panels show the case of unconstrained orbits.
The results of the least square fit and the 1-D, $1\sigma$ scatter
around the best fitting line are indicated in each panel.
}
\label{fig:sph2}
\end{figure}

The amplitude of the random errors is better appreciated 
in fig.~\ref{fig:sph2}, where  we plot
the  Cartesian X-component of  
true {\it vs. } FAM velocities of $\sim 1000$ randomly selected particles.
The parameters of the best linear fit 
and its 1-D, $1\sigma$ scatter are shown in each panel.
The diagonal  line in each panel is plotted to guide the eye.
The two upper panels show the results when the particles are 
forced to move along radial orbits.
When this constraint is removed (lower panels), 
the shot noise error
causes a $120-150 \kms $ uncertainty
in FAM reconstructions which is clearly  visible 
both in x- (bottom-left) and s-space (bottom-right).
The small systematic errors seen in fig.~\ref{fig:sph1} are also visible
in fig.~\ref{fig:sph2}  and cause the FAM velocities
to be slightly underestimated, especially when the
reconstruction is performed in x-space. 
Systematic errors can be quantified by the deviation
of the slope of the best fitting line from unity.
The main source for the systematic mismatch is the softening
used in the TREECODE to compute the force field.  As already pointed
out by NB, the amplitude of FAM velocities decreases when increasing
the softening parameter, along with the scatter around the best
fitting line.  This is true also in the spherical infall experiment,
as we have checked by running the same FAMr with a smaller
softening parameter of $\epsilon=0.25$ Mpc.  As expected, using a smaller
softening reduces  systematic errors and increases
the slope of the best fitting line from $0.92$ to $0.95$.

We have also tested  FAMz with spherical infall in a universe with 
$\Omega_m=0.3$ and $\Omega_{\Lambda}=0.7$.  
FAMz velocities are still free of systematic biases and the 
relative random errors are similar to those found in the previous 
tests in a flat, $\Omega_m=1$ universe. 

\section{The Mock $N$-body Catalogs}
\label{sec:mock}       

In Sections~\ref{sec:Nbodyvol} and \ref{sec:Nbodyflux} we will perform
more realistic and demanding tests of FAM using a suite of mock
catalogs extracted from two $N$-body simulations performed by Cole \etal
(1998).  The initial conditions of the two simulations were generated,
respectively, from two cosmological models of Cold Dark Matter: a flat
$\Lambda$CDM model with $\Omega_m=0.3$ and $\Omega_{\Lambda}=0.7$, and a
flat $\tau$CDM universe with $\Omega_m=1.0$ and power spectrum shape
parameter $\Gamma=0.25$.  Both simulations were ran with an AP$^3$M
code loaded with $192^3$ particles in a box of side $345.6 \hmpc$.  In
both simulations the initial amplitude of the density fluctuations was
normalized to the observed abundance of galaxy clusters. This requires
setting $\sigma_8=0.55 \Omega_m^{-0.6}$ (Eke, Cole \& Frenk 1996), where
$\sigma_8$ is the linear rms mass density fluctuation in top-hat
spheres of radius $8\hmpc$ (which hereafter we indicate with TH8).  
Before generating the mock catalogs 
the distribution of $N$-body particles has been re-sampled by assigning 
a constant probability to each particle which is used to perform 
a Montecarlo rejection procedure from a Poisson distribution with mean equal to the required 
probability. The original$N$-body particles particles are thus substituted by 
0,1,2 or on rare occasions an even higher number of particles
that retain positions and velocities of the parent object.
Particles at the same locations are usually collapsed into 
a single objects with higher mass. Occasionally, however, 
roundoff errors cause them to be 
retained as close pairs or triplets of particles.
Furthermore, we have cooled down the particles' peculiar velocities since they are
significantly larger than those of real galaxies.  The new peculiar
velocities ${\bf v}$ are obtained in two steps (Davis, Nusser \&
Willick 1996).  First we perform a mass-weighted
smoothing of the N-body velocities, ${\bf v}_{\scriptscriptstyle Nbody}$,
with a spherical top hat filter of radius $1.5 \hmpc$ 
and obtain a smoothed velocity
field ${\bf v}_{1.5}$ Then the two vector fields are added linearly to
give ${\bf v}=0.7\times{\bf v}_{\scriptscriptstyle Nbody}+0.3\times{\bf v}_{1.5}$.  The
new pairwise velocity dispersion measured for object with a relative
separation of $1.0 \hmpc$ is $\sim 200 \kms$, comparable to the value
measured by Strauss, Ostriker \& Cen (1998) for galaxies outside high
density region in the Optical Redshift Survey.

From these two simulations we have extracted 4 sets of mock catalogs.
Each catalog lists particle positions, redshifts, and peculiar
velocities in a  spherical region of radius of $80 \hmpc$ centered
on a Local-Group look-alike particle (see Branchini \etal 1999 for details).
Each set contains 5+5 independent mock catalogs, corresponding to the two
simulations, so that in total we have 40 catalogs.  The four
sets are termed VL, VL5TH, FL, and FL5TH and their main properties are
summarized in table~\ref{tab:mock}.  The sets VL and VL5TH  are 
volume-limited catalogs, while FL and FL5TH  are flux-limited.

\begin{figure}
\vspace{13truecm}
{\includegraphics{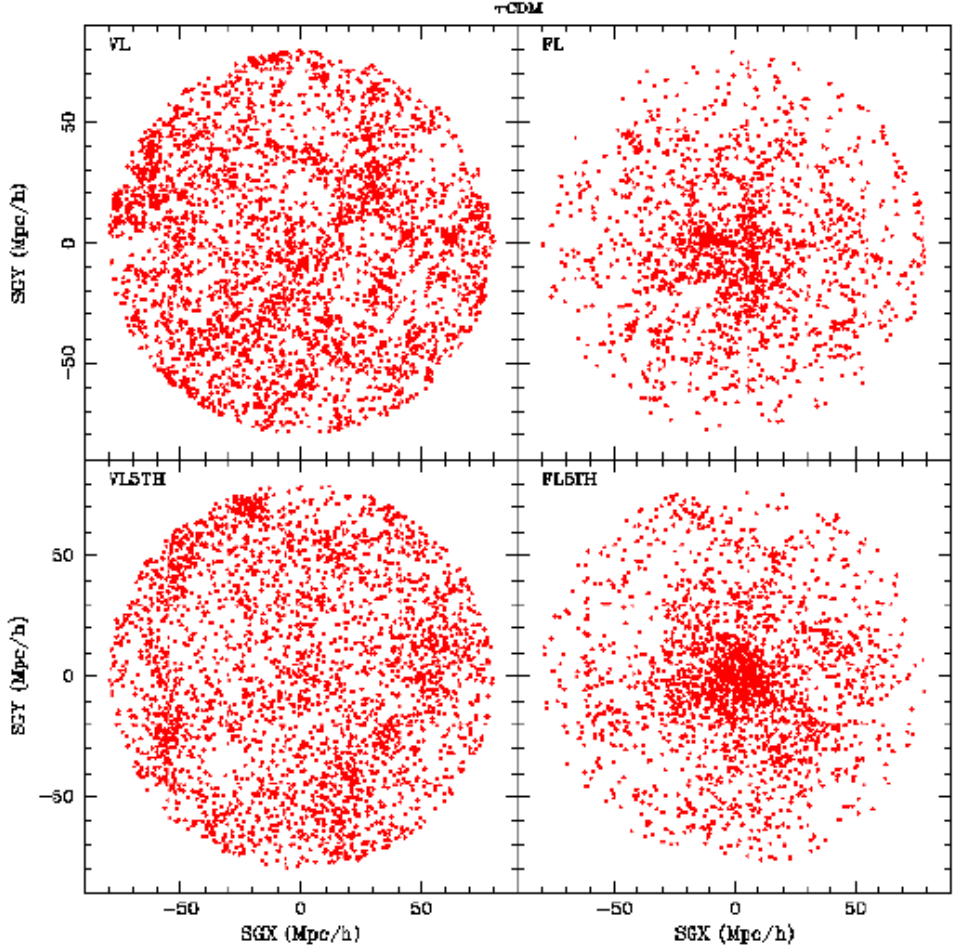}}
\caption{ Particle distribution in x-space in slices of thickness $20
\hmpc$ cut through mock catalogs in the
$\tau$CDM $N$-body simulation.  Each panel shows a VL (top-left), VL5TH
(bottom-left), FL (top-right), and FL5TH (bottom-right) mock
catalog selected from the 4 sets.}
\label{fig:maps}
\end{figure}

A catalog in the VL set is generated by randomly extracting
20,000  particles in a sphere of radius $80 \hmpc$, half of
which sample the inner sphere of radius $40\hmpc$ and the rest are
contained in the shell between $40$ and $80\hmpc$.
Different masses have
been assigned to internal and external particles to guarantee the same
average mass density throughout the volume.  A typical VL mock catalog
is displayed in the top-left panel of fig.~\ref{fig:maps} which shows
a slice of thickness $20 \hmpc$ cut through a VL mock catalog extracted
from the $\tau$CDM simulation.

To obtain the VL5TH catalogs we first smooth in x-space
the particles' distribution in the $N$-body simulation
in a mass-weighted fashion
with a spherical top-hat filter of radius $5 \hmpc$ and then we
Poisson sample the smoothed density field with 20,000 particles, half of which
inside $40\hmpc$.  
The bottom-left panel of fig.~\ref{fig:maps} shows
the x-space particle distribution in one of the VL5TH catalogs.  The particle
distribution appears to be less clustered than in the unsmoothed case,
as expected. 
The redshifts of mock galaxies are obtained by adding the
radial component of the 5TH-smoothed $N$-body velocities 
to the particles' distance (expressed in $\kms$).

The  sets FL and FL5TH are designed to mimic 
 the PSC$z$ flux-limited survey.
A FL  mock catalog typically contains
$\sim 8000$ particles  within a sphere of $80 \hmpc$,  extracted 
from the $N$-body simulation according to the selection 
function of PSC$z$ galaxies as determined by Branchini \etal (1999):
\begin{equation}
 \phi(r)= \left\{\begin{array}{ll}
   1& \mbox{if $r\le 6\hmpc$}  \\
  \left(\frac{r}{r_0}\right)^{-1.08}\left(1+ {{r}^2\over{r_{\star}^2}} \right)^{-1.83} 
&\mbox{if $r>6\hmpc$} 
\end{array}\right.
\label{eq:self}
\end{equation}
where $r$ is the distance from the center of the sphere in $\hmpc$, $r_0=
6.05\hmpc$, and
$r_{\star}=87 \hmpc$.  
The mass assigned to each galaxy  is
equal to the inverse of the selection 
function at the galaxy position $\phi(\vx)$.
An example of 
catalog in the FL set is shown in the top-right panel of fig.~\ref{fig:maps}. The
particle distribution is concentrated towards the center due to the
cut in flux.  To obtain the FL5TH catalogs we follow a two step
procedures.  First we perform a mass-weighted smoothing of the particle distribution
in the x-space within each of the FL mock catalogs using a top-hat filter
with an adaptive radius of $R_{TH}={\rm Max}[5,l] \hmpc$, where $l$ is
the average inter-particle separation at the generic position.
This smoothing radius compromises the need of avoiding  
nonlinear effects with that of minimizing shot noise errors. 
Then we sample the  smoothed density  with $\sim
8000$ particles with a radial distribution according to the 
selection function (eq.~\ref{eq:self}). The particle distribution in one of the FL5TH mock
catalogs is shown in the bottom-right panel of fig.~\ref{fig:maps}.
As for the VL5TH case, the redshifts  of the mock
galaxies in the  FL5TH catalogs are obtained using the 
velocities smoothed with a top hat filter
on the variable scale  $R_{TH}$.

\begin{table}
\centering
\caption[]{Main properties of the four sets of mock catalogs.
Name: Label of the set.  \#$_{Cat}$: Number of catalogs in each set
(representative of $\Lambda$CDM and  $\tau$CDM cosmological models).
 \#$_{Part}$: Number of particles in each catalog. Sampling:
Monte Carlo sampling technique adopted.   R$_{TH}$: radius of the top-hat filter
used to smooth the original simulation ($l$ represents the average 
interparticle separation). R$_{max}$: external radius of each sample.
}
\tabcolsep 2pt
\begin{tabular}{cccccc} \\  \hline
Name &  \#$_{Cat}$  &  \#$_{Part}$ & Sampling & R$_{TH}$ ($\hmpc$)& R$_{max}$ ($\hmpc$) \\ \hline 
VL & 5+5 & 20000 & 10000+10000 & $0$ &$80$ \\ 
VL5TH & 5+5 & 20000 & 10000+10000 & $5$&$80$   \\ 
FL & 5+5 & $\sim 8000$ & PSC$z$ Selection & $0$ &$80$  \\ 
FL5TH & 5+5 & $\sim 8000$ & PSC$z$ Selection & Max $\left[ 5,l \right]$ &$80$ \\ \hline
\end{tabular}
\label{tab:mock}
\end{table}

\section{Tests with Volume-Limited Catalogs}
\label{sec:Nbodyvol}

In this Section we apply  FAM to the suite of 
VL and VL5TH  catalogs.
Unless otherwise stated, all FAM reconstructions were performed with a
softening  $\epsilon=0.5 \hmpc$, a tolerance  $tol=10^{-6}$, and
${n_{max}}=6$ basis functions. 
Decreasing $tol$ or increasing
${n_{max}}$ makes little change to the final results.

\subsection{Unsmoothed Volume-Limited: VL}
\label{sec:volu}

\begin{figure}
\vspace{13truecm}
{\includegraphics{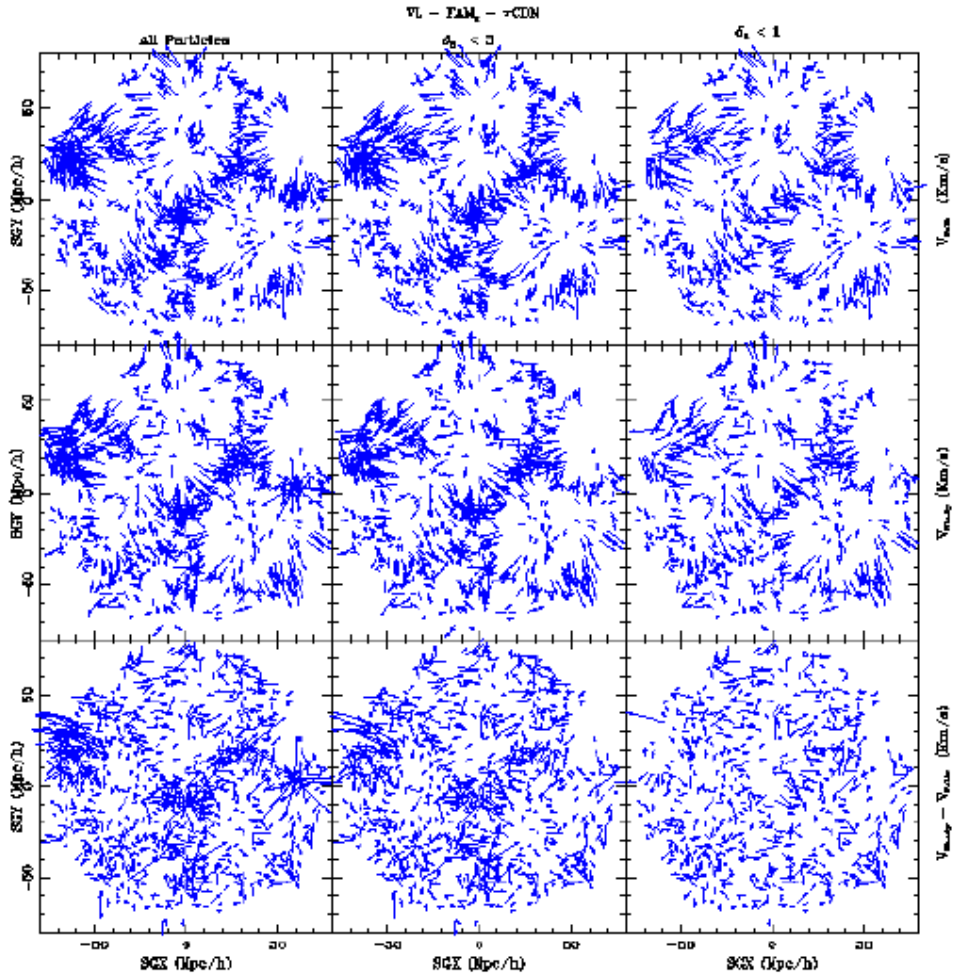}}
\caption{Maps of 2D-projected peculiar velocities for points residing
in a slice of thickness $6 \hmpc$ cut through a VL-$\tau$CDM
catalog.
The length of the vectors is drawn in units of $1 \hmpc=50 \kms$.
The top row shows the FAMz-predicted velocities. $N$-body
velocities are shown in the middle row. The velocity residuals, ${\bf
v}_{\scriptscriptstyle Nbody}-{\bf v}_{\scriptscriptstyle FAMz}$, 
are displayed on the bottom.  The maps
shown in the panels to the left hand side refer to all the
points in the slice while only the velocities of points with $\delta_5
< 3$ and $\delta_5 < 1$ are plotted in the central and right
columns, respectively.  }
\label{fig:vmap}
\end{figure}

In fig.~\ref{fig:vmap} we show maps of the $N$-body and FAMz
peculiar velocities for one of the VL catalogs.  The
vectors represent the 2D-projected peculiar velocities of particles in
a slice of thickness $6 \hmpc$ cut through one of the VL-$\tau$CDM
catalogs.  The length of the vectors is drawn in units of $1 \hmpc=50 \kms$.
The top-left panel shows FAMz 
velocities for all the particles in the slice. The $N$-body velocities of the same
particles are shown in the middle-left panel, and the velocity residuals, ${\bf
v}_{\scriptscriptstyle Nbody }-{\bf v}_{\scriptscriptstyle FAMz }$,   
in the the bottom-left. Points with the largest
residuals  are located in regions of high density.  These regions are
characterized by  virial velocities  which
are not  modeled correctly by FAM.  In these regions, FAM typically predicts a
large, coherent infall into the gravitational potential wells of
density peaks.  One example is the region centered around the point
with $({\rm SGX,SGY})\approx(60,0)$. The signature of virial motions is
clearly visible in the $N$-body velocity map but is completely absent
in FAM.  Instead, FAM predicts coherent inward streaming velocities.

FAM is meant to model gravitational dynamics in the mildly nonlinear
regime.  So when we move away from high density environments we expect
an improvement in the agreement between the $N$-body and FAM
velocities.  To verify this hypothesis, we define $\delta_5$ as the density
contrast smoothed with a TH5 filter
and we only compare   velocities of  particles in
regions with $\delta_5$ less than a certain value.  The central column
of fig.~\ref{fig:vmap} shows peculiar velocities of particles in
regions with $\delta_5<3$.  Cutting at this density threshold excludes
only $\sim 5$ \% of the particles, but considerably improves the
agreement between the two fields (the ``hot spot'' located at 
$({\rm SGX,SGY})\approx(60,0)$
disappears from the map of the residuals). The agreement further
improves when we exclude points in regions with $\delta_5 \ge 1$
($\sim 20\%$ of the points), as shown by the third row of panels in
fig.~\ref{fig:vmap}.  In this last case the velocity residuals
(bottom-right) are further reduced and are not concentrated around
regions of enhanced density.  Very few large residuals still exist for
points with large $N$-body velocities.
 
\begin{figure}
\vspace{13truecm}
{\includegraphics{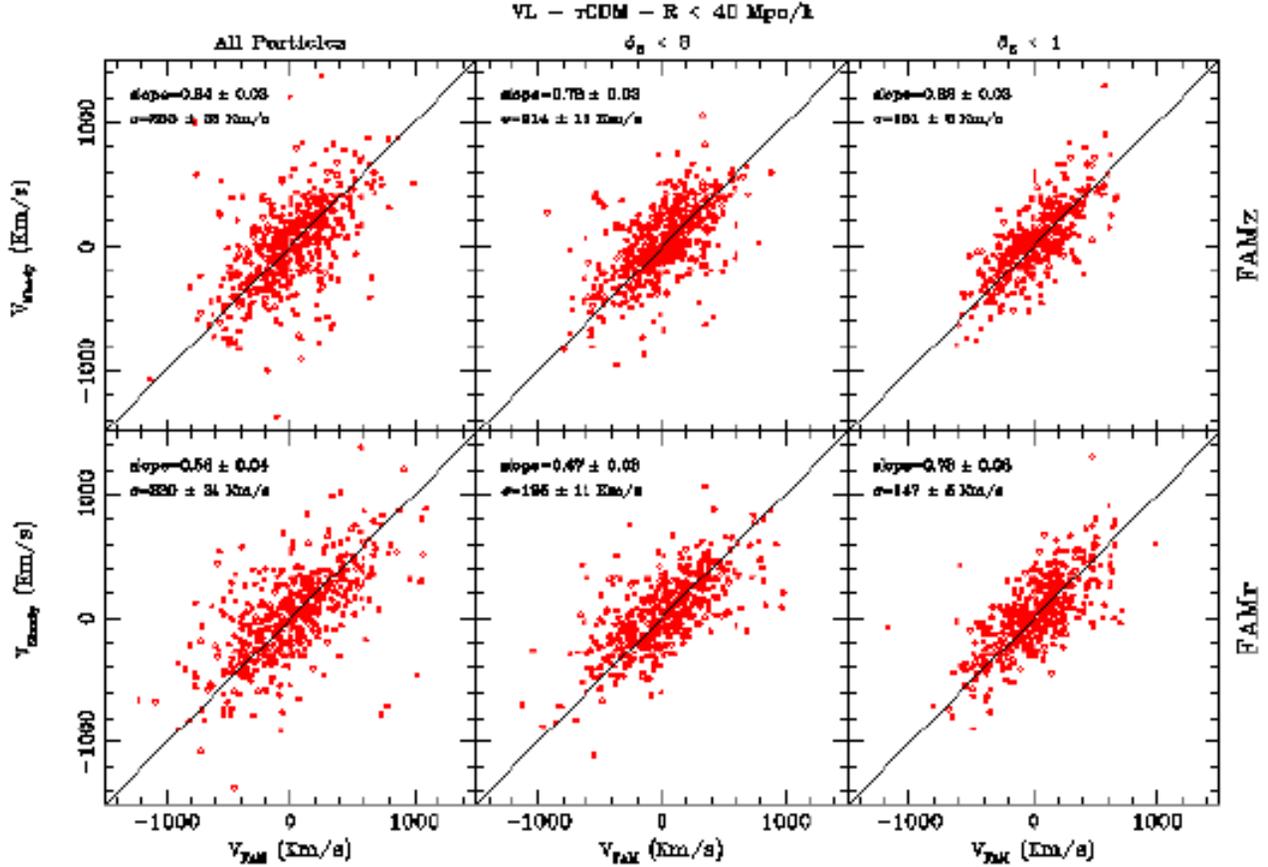}}
\caption{ $N$-body vs. FAM particle velocity scatter plots in x-space 
(bottom panels) and s-space (top panels). 
Only the X-Cartesian velocity compontent of 500 objects, 
randomly selected among 
the 50,000 contained within $40 \hmpc$ from the center of the 5 VL - $\tau$CDM 
catalogs are plotted.
Particles in the left panels are selected from all density environments while only particles 
with $\delta_5 < 3$
and $\delta_5 < 1$ are displayed in the central and right panels, respectively. 
In all the plots, we show the slope and the 1-D velocity scatter around the fit (which uses all 
the  50,000 objects available) while
the straight line shows the expected correlation.
}
\label{fig:vv_FAMz_v}
\end{figure}

Fig.~\ref{fig:vv_FAMz_v} shows a more quantitative point by point
{\it v-v} comparison.  In each panel the Cartesian X-components of the
 $N$-body velocities is plotted  against the corresponding
FAMz (top panels) and FAMr (bottom panels) peculiar velocities.  
Each panel shows the velocities of 
500 particles randomly selected from the  total of 50,000
lying within a distance of $ 40 \hmpc$  in all  the 5 VL-$\tau$CDM catalogs.
The left
column shows the velocities of all particles. The
panels in the middle and right columns show only the velocities of
particles with $\delta_5 < 3$ and $\delta_5 < 1$, respectively.

To  quantify the match between
FAM and $N$-body velocities we perform a linear regression of
the X-Cartesian components of
$N$-body velocities on FAM assuming errors on the $N$-body velocities only,
using all the particles 50,000 in the 5 VL-$\tau$CDM  catalogs. 
The average slope and 1-D scatter of the regression of the 5
catalogs in each set are shown in the corresponding panels of
fig.~\ref{fig:vv_FAMz_v} along with the rms values
of the average slope and 1-D velocity scatter over the 5 catalogs.
The reasoning behind performing the linear regression 
of $N$-body velocities on FAM with errors on the 
$N$-body velocities is twofold.
First, as pointed out by Berlind, Narayanan and Weinberg (2000),
$N$-body velocities consist of a large-scale contribution predicted from
the mass distribution plus an uncorrelated small-scale contribution
that can be regarded as a thermal noise. This suggests putting all the
errors on  the $N$-body velocities, while leaving the FAM-reconstructed 
velocities on the X-axis.
Second, this choice mimics the real velocity-velocity comparisons
in which the observed ($N$-body) velocities are compared with
theoretical (FAM) predictions and the regression is performed
putting all the errors in the observed quantity.

The top and bottom panels on the left hand side of
fig.~\ref{fig:vv_FAMz_v} show a clear correlation between 
$N$-body and FAM velocities with  small scatter around the best fitting line,
both in x- and s-space, although in both cases FAM 
overestimates the amplitude of the velocities.
A few out-liers are present in the plots.   
However, while FAMz (top-left) underestimates the amplitude of
the velocities of the out-liers, FAMr (bottom-left) does the
opposite. So the slopes of the linear fits in FAMr are smaller than
the corresponding slopes in FAMz.  
When the comparison is restricted
to particles with $\delta_5 < 3$ (central row) the number of
out-liers decreases and the agreement with $N$-body velocities improves
significantly, indicating that discrepant velocities arise in high
density regions as suggested by the velocity maps in
fig.~\ref{fig:vmap}. This trend is confirmed by the scatter plots on
the right column which only include points with $\delta_5 < 1$,
well outside high density environments.  
In this last case the slopes of the FAMr
and FAMz best fits are quite similar and even closer to unity.  Furthermore,
the out-liers have almost disappeared and the 1-D velocity scatter around the fit
decreased to a very small level of $\approx 150 \kms$, to be compared 
with a  typical 1-D rms velocity dispersion of $\approx 350 \kms$
for the objects in the VL catalogs.  
We conclude that FAMz and FAMr perform similarly outside high density regions (i.e. for
$\sim 95 \%$ of the points).
The results of the fits shown in the plots are summarized in the upper
part of table~\ref{tab:tcdm}.
The column labeled by  $4\sigma$, indicates the result of linear regressions
in which  out-liers in the {\it v-v} scatter plots are
eliminated by iteratively discarding
all points deviating by more than $4\sigma$ from the best fitting
line.  The $4\sigma$-clipping procedure converges in a few iterations.
The final outcome of this procedure is similar to 
that of imposing a density
cut at $\delta_5=1$  and returns a best fit closer to unity.

So far only particles within $40 \hmpc$ were included in the
{\it v-v} comparisons, although all particles out to
$80\hmpc$ were used in  FAM reconstruction. This is to guarantee
dense sampling (the outer part of these catalogs is more sparsely
sampled) and to minimize the effect of the external tidal field.
Including all particles out  to $80 \hmpc$ does not affect the
result appreciably, apart from a systematic $\sim 5 \%$ increase in
the slope of the best fitting line, which probably quantifies the bias
introduced by having assumed a homogeneous mass distribution beyond
$40 \hmpc$ in the former experiments.  Similar considerations apply to
the softening parameter. While, as shown in the NB analysis,
increasing the softening parameter decreases the amplitude of
FAM velocities, decreasing it from $0.5$ to $0.25 \hmpc$ leaves the
results of the {\it v-v} comparison unchanged.

We have also checked the dependence of our results on the
cosmological model by repeating the same analysis with the 5 
VL catalogs extracted from the $\Lambda$CDM $N$-body
experiment. The parameters of the linear fits to the {\it v-v}
comparisons are also listed in  table~\ref{tab:tcdm}
and are similar to those found in the $\tau$CDM scenario.  As for
the $\tau$CDM case, FAMr velocities are systematically larger than
the FAMz ones and the mismatch with $N$-body velocities
decreases when moving away from high
density regions.  The effect is more dramatic in the $\Lambda$CDM universe,
as it can be verified by comparing the results of the fits
with no density cut with those of the $\delta_5=3$ cut.  The reason
for this behavior is that the $\Lambda$CDM model is characterized by a
larger value of $\sigma_8$, i.e. by a larger number of regions with
high overdensity where FAM fails to predict the correct peculiar
velocities.

\begin{table}
\centering
\caption[]{FAMr and FAMz results for the VL (upper part) 
and FL (lower part) experiments for two different cosmological models
($\tau$CDM and $\Lambda$CDM).
The average slope,  $b$, and 1-D velocity scatter  $\sigma$ (in $\kms$) are shown along with 
their rms scatter, measured over the 5 independent experiments.
The best fit values are listed for various overdensity thresholds $\delta_5$
and for the case of $4\sigma$-clipping. } 
\tabcolsep 2pt
\begin{tabular}{ccccccccccc} \\  \hline
& & & FAMr   &  &  &  &  FAMz & & & \\ \hline 
Model &   & All            & $\delta_5 < 3 $ & $\delta_5 < 1$ & $4\sigma$ &  All           & $\delta_5 < 3$  & $\delta_5 < 1 $ & $4\sigma$ \\ \hline
VL-$\tau$CDM & $b$ & $0.58\pm 0.04$ & $0.67 \pm 0.03$  & $0.78 \pm 0.03$  & $0.80 \pm 0.03$ & $0.64\pm 0.03$ & $0.78 \pm 0.03$ & $0.82 \pm 0.03$ &$0.82 \pm 0.04$ \\
& $\sigma$& $230 \pm 24$ & $195 \pm 11$ & $147 \pm 5$ &  $131 \pm 5$ & $253\pm 33$ & $214 \pm 15$ & $161 \pm 8$ & $141 \pm 8$  \\ \hline
VL-$\Lambda$CDM & $b$ & $0.43\pm 0.03$ & $0.71 \pm 0.03$ & $0.84 \pm 0.04$ & $0.88 \pm 0.03$ & $0.68\pm 0.02$ & $0.82 \pm 0.03$ & $0.86 \pm 0.04$& $0.91 \pm 0.06$  \\
& $\sigma$& $287\pm 55$ & $161 \pm 12$ & $124 \pm 7$ & $105 \pm 10$ & $ 295 \pm 59$ & $167 \pm 12$ & $126 \pm 6$ & $108 \pm 6$ \\ \hline \hline
FL-$\tau$CDM & $b$ & $0.67\pm 0.03$ & $0.72 \pm 0.03$  & $0.78 \pm 0.02$ & $0.81 \pm 0.03$ & $0.74\pm 0.04$ & $0.80 \pm 0.04$ & $0.88 \pm 0.03$ & $0.92 \pm 0.05$ \\
 & $\sigma$& $227\pm 12$ & $201 \pm 10$ & $171 \pm 12$ & $152 \pm 14$ & $242\pm 16$ & $212 \pm 14$ & $170 \pm 13$ & $145 \pm 12$  \\ \hline
FL-$\Lambda$CDM & $b$ & $0.51\pm 0.02$ & $0.70 \pm 0.03$  & $0.86 \pm 0.03$ & $0.94 \pm 0.03$ & $0.67\pm 0.06$ & $0.86 \pm 0.04$ & $0.93 \pm 0.02$ & $0.96 \pm 0.03$ \\
 & $\sigma$& $223\pm 26$ & $168 \pm 24$ & $135 \pm 14$ & $108 \pm 9$ & $ 240 \pm 60$ & $164 \pm 24$ & $126 \pm 15$ & $99 \pm 11$ \\ \hline
\end{tabular}
\label{tab:tcdm}
\end{table} 

\begin{figure}
\vspace{13truecm}
{\includegraphics{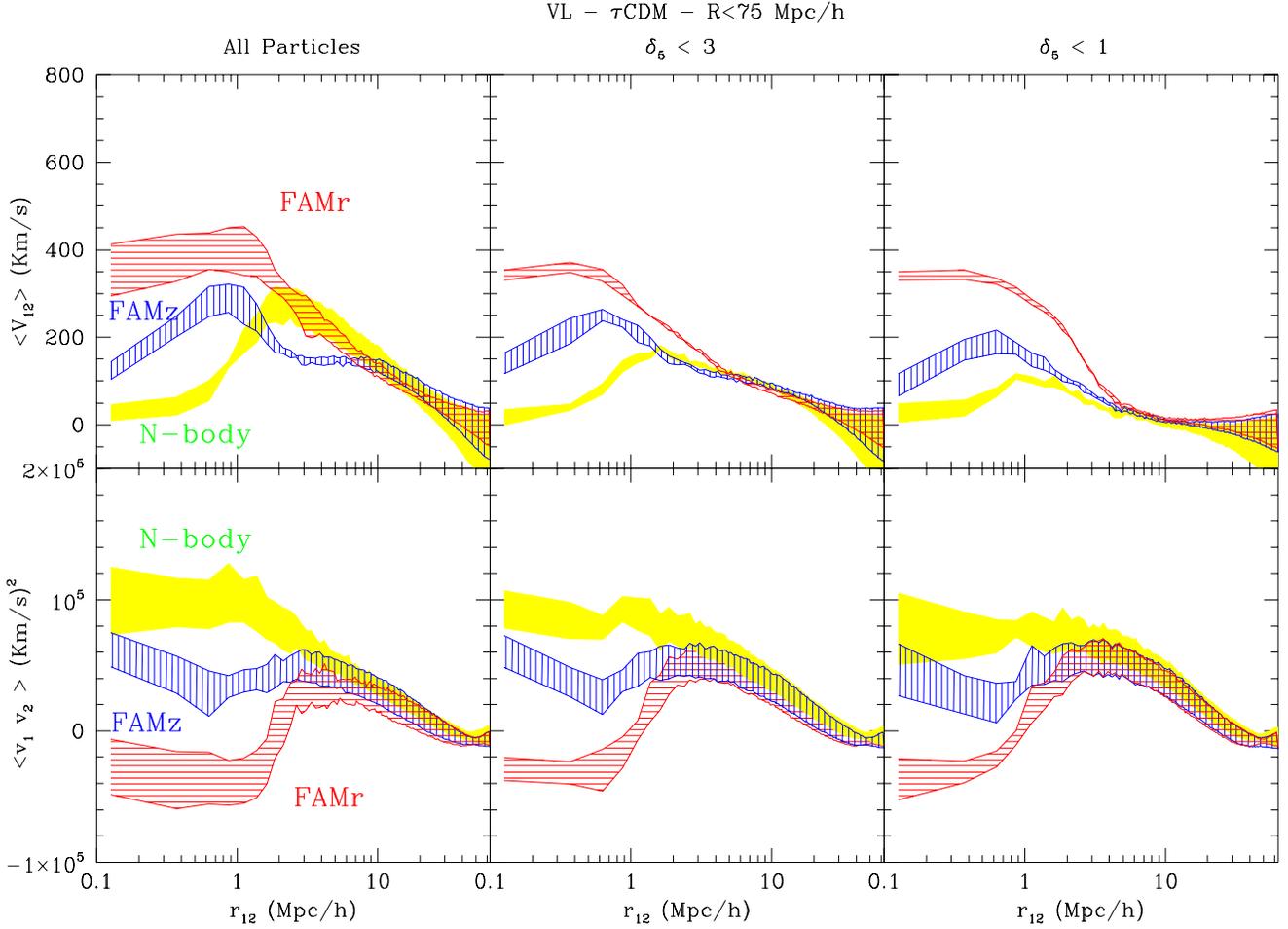}}
\caption{Relative pairwise velocity $ \langle  V_{12} \rangle$ (upper panels) 
and velocity correlation function projected along the separation of the pair 
$ \langle  V_1 V_2 \rangle$ (lower panels), 
as a function of the separation $r_{12}$ averaged over the 5 VL-$\tau$CDM  
catalogs.
The ``ribbons'' represent the $1\sigma$ uncertainty intervals around the mean value.
Vertically and horizontally-dashed ribbons show the results of the FAMz and FAMr,
respectively.
The dark strips refer to the $N$-body experiment. 
The plots in the central row and those in the row on the right have been 
obtained by considering  points with with $\delta_5 < 3 $ and $\delta_5 < 1$
and  within $75 \hmpc$. 
All the particles available have been used in the plots on the left.
}
\label{fig:vcorr_FAMz_v}
\end{figure}

The {\it v-v} scatter plots tell us little about the ability of FAM at
recovering the correlation properties of the velocity field. These
have been investigated using two different statistical tools.  The
first is the average relative pairwise velocity $\langle V_{12}
\rangle \equiv \langle [\vv(\vx+\vr_{12}) - \vv(\vx)]\cdot\ve_{12}
\rangle $, where the averaging is over all pairs of objects at
separation $|\vr_{12}|$, and $\ve_{12}$ is a unit vector along the
the separation defined so that
approaching pairs have $\langle V_{12} \rangle>0$.
The second statistics is the {\it v-v} correlation function
projected along $\vr_{12}$, defined as 
$\langle V_1 V_2 \rangle \equiv
\langle \vv(\vx) \cdot \ve_{12} \vv(\vx+\vr_{12}) \cdot \ve_{12} \rangle $
(Gorski 1988).  These statistics have been computed for each of the VL
catalogs.  The results are shown in fig.~\ref{fig:vcorr_FAMz_v} for
the $\tau$CDM case.  Each ``ribbon'' represents the $1\sigma$
uncertainty interval around the mean values of $\langle V_{12} \rangle$ and
$\langle V_1 V_2 \rangle$, averaged over the 5 reconstructions and plotted as
a function of $|\vr_{12}|$.  The vertically-dashed and 
horizontally-dashed strips show the results of the FAMz and FAMr experiment. The
dark strips show the ``true'' statistics computed from the $N$-body velocities.
Similarly to fig.~\ref{fig:vcorr_FAMz_v}, the two plots to the right
have been obtained using all the particles within $75 \hmpc$ while
only particles with $\delta_5 < 3 $ and $\delta_5 < 1$ have been used
in the two other plots.

The top-left panel shows that at separations smaller than $\sim 1
\hmpc$, both  $\langle V_{12} \rangle_x$ and $\langle V_{12} \rangle_s$ are
systematically larger than the $N$-body result.  This is the signature
of what we have already spotted in the velocity maps, i.e. that in
high density region (where most of the pairs with small relative
separation reside) FAM predicts coherent infall ($\langle V_{12}
\rangle >0$) instead of disordered thermal motions ($\langle V_{12}
\rangle =0$).  
Also, the FAMr-predicted infall velocities are larger
that the FAMz ones at separation smaller that few Megaparsecs.
As we have already noticed in  the linear fit to
the {\it v-v} comparisons,  this systematic  discrepancy
is caused by the smearing of density peaks in s-space
which decrease the number of pairs having one member in over-dense regions
and suppress the infall signal in the FAMz reconstructions.
The same considerations apply to the plot
on the bottom-left corner, which shows that the spurious infall makes
$\langle V_1 V_2 \rangle$ of FAM velocities  systematically smaller
than the $N$-body one up to scales of $\sim 3 \hmpc$.  
These considerations are verified by the middle and right columns  of 
fig.~\ref{fig:vcorr_FAMz_v}.  Removing the high density
regions does not change significantly the FAM average results but only
the scatter. In contrast, the $N$-body results change appreciably,
especially at small separations. This indicates that most of the
infall signal in the FAM velocities  comes from close pairs (since the
solution of first approach is the one preferentially found at the
minimum of the action) which do not necessarily reside in high density
environments.  For the $N$-body experiment, however, the strong infall
signature around $5 \hmpc$ disappears when neglecting high density
regions, meaning that most of the infall signal is contributed by pairs
in which at least one member is located in a high density spot.

\subsection{Smoothed Volume-Limited: VL5TH}
\label{sec:vols}

The analysis  in Section~\ref{sec:volu} revealed that
when high density regions are excluded, FAM  predicts an almost unbiased
peculiar velocity field and reproduces its correlation properties
on scales larger than a few Mpc.
Instead of discarding regions
where the velocity field is highly non-linear, we can apply FAM on 
a particle distribution obtained by sampling a smoothed version 
of the original particle distribution. Clearly, this smoothing 
operation is rather unrealistic, since real peculiar velocities 
are too sparse to be smoothed. Thus, the test performed in this Section  
is ideal. The aim is to minimize ``thermal noise'' and to
concentrate on the effects of sampling errors on FAM velocity prediction.
Therefore, here we repeat the analysis of Section~\ref{sec:volu} 
on the 5TH-smoothed, volume limited mock catalogs VL5TH.
However, since the thermal noise has been eliminated by the 
smoothing procedure, in this section we assume that 
errors only affect FAM velocity prediction and 
perform linear regressions 
of  FAM velocities on $N$-body, with errors on the Y-axis 
only.

\begin{figure}
\vspace{13truecm}
{\includegraphics{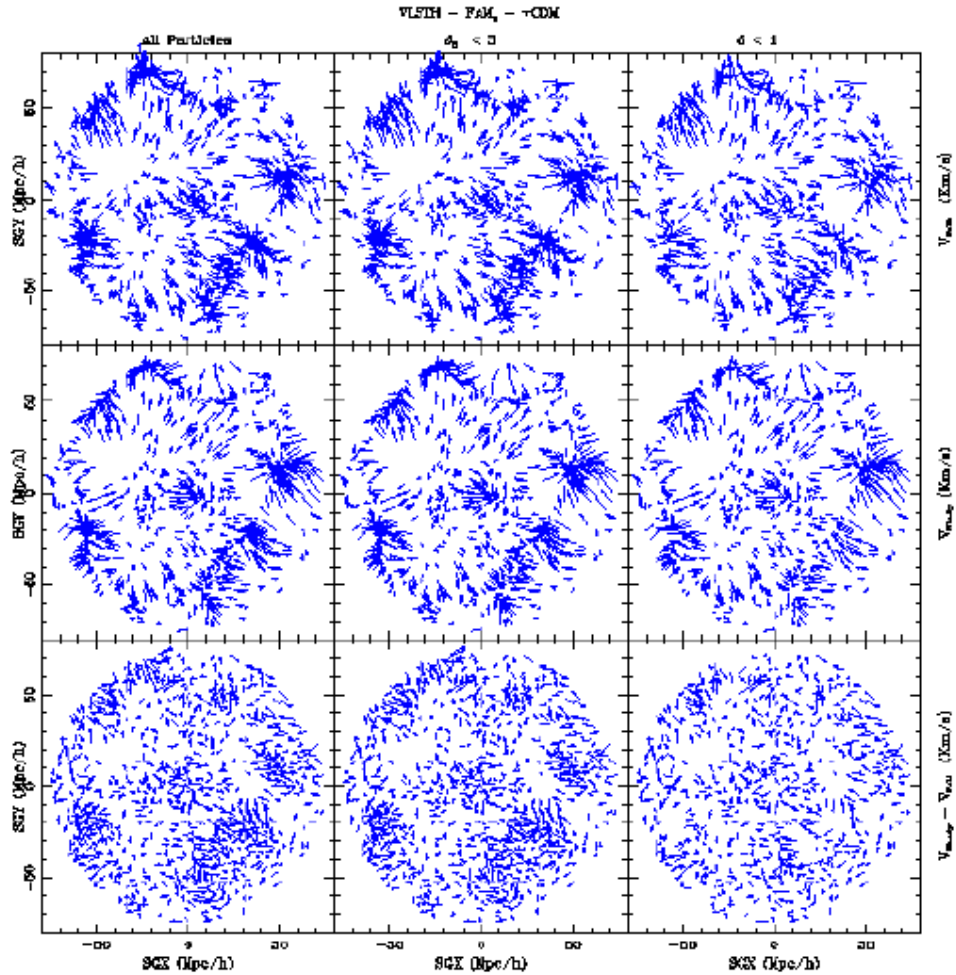}}
\caption{Peculiar velocities in a 
slice of thickness $6 \hmpc$ cut through one of the VL5TH-$\tau$CDM catalogs.
Plots in the top row: FAMz velocities. 
Plots in the middle row: 5TH-smoothed $N$-body velocities.
Plots in the bottom row:  velocity residuals ${\bf v}_{\scriptscriptstyle Nbody}-
{\bf v}_{\scriptscriptstyle FAMz}$. 
Panels on the left: all the points in the slice are shown.
Panels in the middle: points with $\delta_5 < 3$.
Panels to the right: points with $\delta_5 < 1$.
}
\label{fig:vmap_v5TH}
\end{figure}

The plots in the central row of fig.~\ref{fig:vmap_v5TH} show the 
map of the 5TH-smoothed $N$-body peculiar velocities interpolated at the 
particle positions in one of the mock catalogs, for points
selected  in slices similar to those shown in fig.~\ref{fig:vmap}
at the same three different density thresholds.  The
5TH-smoothing operation has erased nonlinear signatures such as the
virial motions in high density regions. The $N$-body flow is dominated
by large, coherent motions which are well reproduced by FAMz (upper
row), as confirmed by the small, incoherent residuals displayed in the
plots on the bottom. 
As expected,  none of the results in this Section change when applying
different density-cuts. Therefore we will only show and discuss the results relative 
to the full sample.

\begin{figure}
\vspace{13truecm}
{\includegraphics{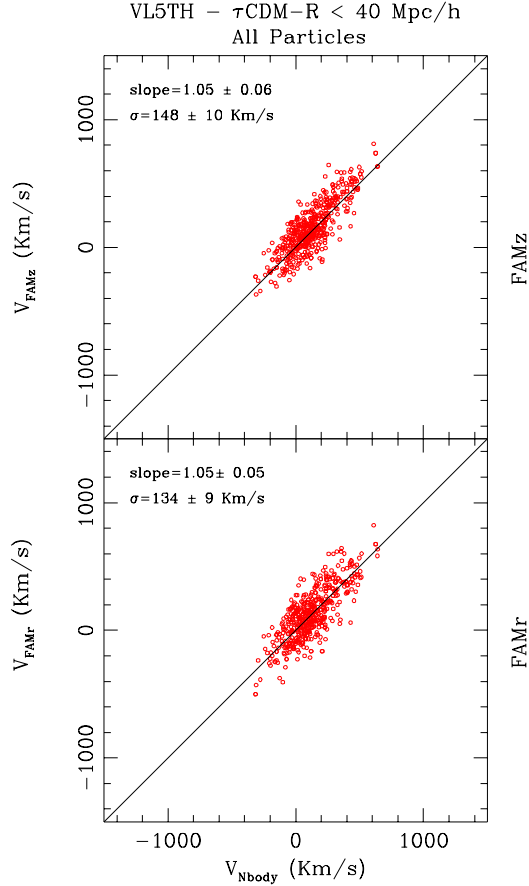}}
\caption{FAM vs. $N$-body velocity scatter plots in x- (bottom panel)
and s-space (top panel) for 500 objects randomly selected from the 5
VL5TH-$\tau$CDM catalogs.  The straight line, showing the expected
correlation, is plotted to guide the eye.  The results of the linear
fit are also displayed.  }
\label{fig:vv_FAMz_v5TH}
\end{figure}

The scatter plots in fig.~\ref{fig:vv_FAMz_v5TH}
show the {\it v-v} comparisons for the 5 VL5TH-$\tau$CDM experiments.
The parameters of the linear  fits are shown in each panel of 
fig.~\ref{fig:vv_FAMz_v5TH} and listed in table~\ref{tab:tcdms}.
The results of the FAMr experiments are very similar to the FAMz ones and do 
not depend on the cosmological scenario.
In all experiments FAM velocities are slightly larger than the 
$N$-body ones. This discrepancy is due to the fact that
the softening parameter in use ($\epsilon=0.5 \hmpc$) is
is much smaller than the smoothing radius.
Increasing the softening to $\epsilon=3.0 \hmpc$ brings FAM 
and $N$-body velocities into agreement, as indicated in 
table~\ref{tab:tcdms}.
The 1-D velocity scatter around the fit is 
significantly smaller than the 1-D rms velocity
dispersion in the VL5TH catalogs ($\approx 240 \kms$).

\begin{table}
\centering
\caption[]{
FAMr and FAMz results for the VL5TH experiments in the
$\tau$CDM (upper part) and $\Lambda$CDM (lower part) cosmological
models.  The average slope, $b$,  and 1-D velocity scatter, $\sigma$, 
along with their rms scatter,
are shown for FAM experiments performed with three different
value of the softening parameter $\epsilon=0.25, \ 0.5, \ 3 \
\hmpc$.}  \tabcolsep 2pt
\begin{tabular}{ccccccccc} \\  \hline
& & & FAMr   &  &  &   FAMz & & \\ \hline 
Model &  & $\epsilon=0.25$ &  $\epsilon=0.5$ & $\epsilon=3.0$ &$\epsilon=0.25$ & $\epsilon=0.5$ &  $\epsilon=3.0$ \\ \hline
VL5TH-$\tau$CDM & $b$ & $1.07\pm 0.08$ & $1.05 \pm 0.05$  & $1.08 \pm 0.10$  &  $1.08\pm 0.09$ & $1.05 \pm 0.06$ & $1.02 \pm 0.09$ \\
& $\sigma$& $154\pm 22$ & $134 \pm 9$ & $112 \pm 8$ &  $147\pm 13$ & $148 \pm 10$ & $118 \pm 8$ \\ \hline
VL5TH-$\Lambda$CDM & $b$ & $0.93\pm 0.06$ & $1.03 \pm 0.05$  & $0.94 \pm 0.06$  &  $0.93\pm 0.10$ & $1.05 \pm 0.06$ & $0.94 \pm 0.10$ \\
 & $\sigma$& $106\pm 21$ & $92 \pm 3$ & $97 \pm 21$ &  $ 115 \pm 24$ & $98 \pm 6$ & $101 \pm 24$ \\ \hline
\end{tabular}
\label{tab:tcdms}
\end{table}

\begin{figure}
\vspace{13truecm}
{\includegraphics{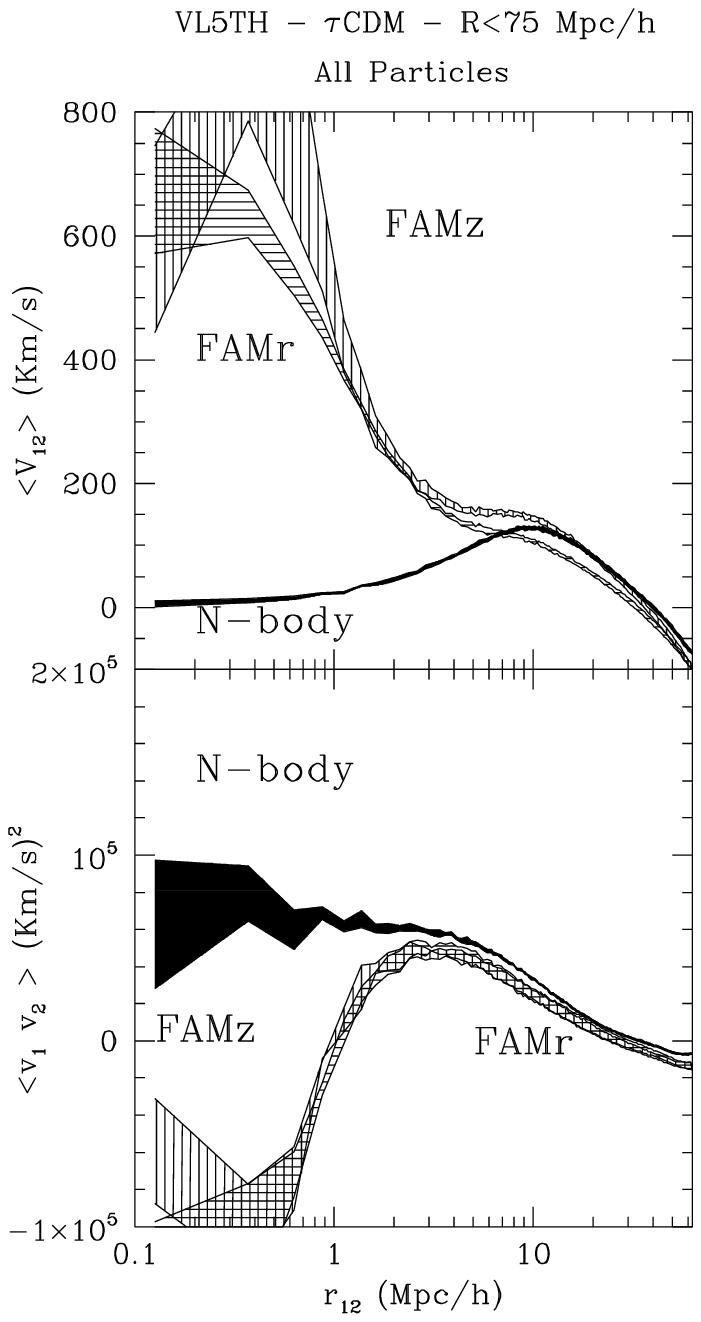}}
\caption{Relative pairwise velocity $\langle V_{12}
\rangle$ (upper panel) and velocity
correlation function projected along the pair separation, 
 $\langle V_1 V_2 \rangle$ (lower panel), computed from
the 5 VL5TH-$\tau$CDM catalogs. The different strips represent
$1\sigma$ uncertainty intervals around the mean values.  Vertically
and horizontally-dashed strips refer to FAMz and FAMr
experiments, respectively. The dark ``ribbon'' refers to the $N$-body
case.  The correlations are computed using all the particles within
$75 \hmpc$.  }
\label{fig:vcorr_FAMz_v5TH}
\end{figure}

\begin{figure}
\vspace{13truecm}
{\includegraphics{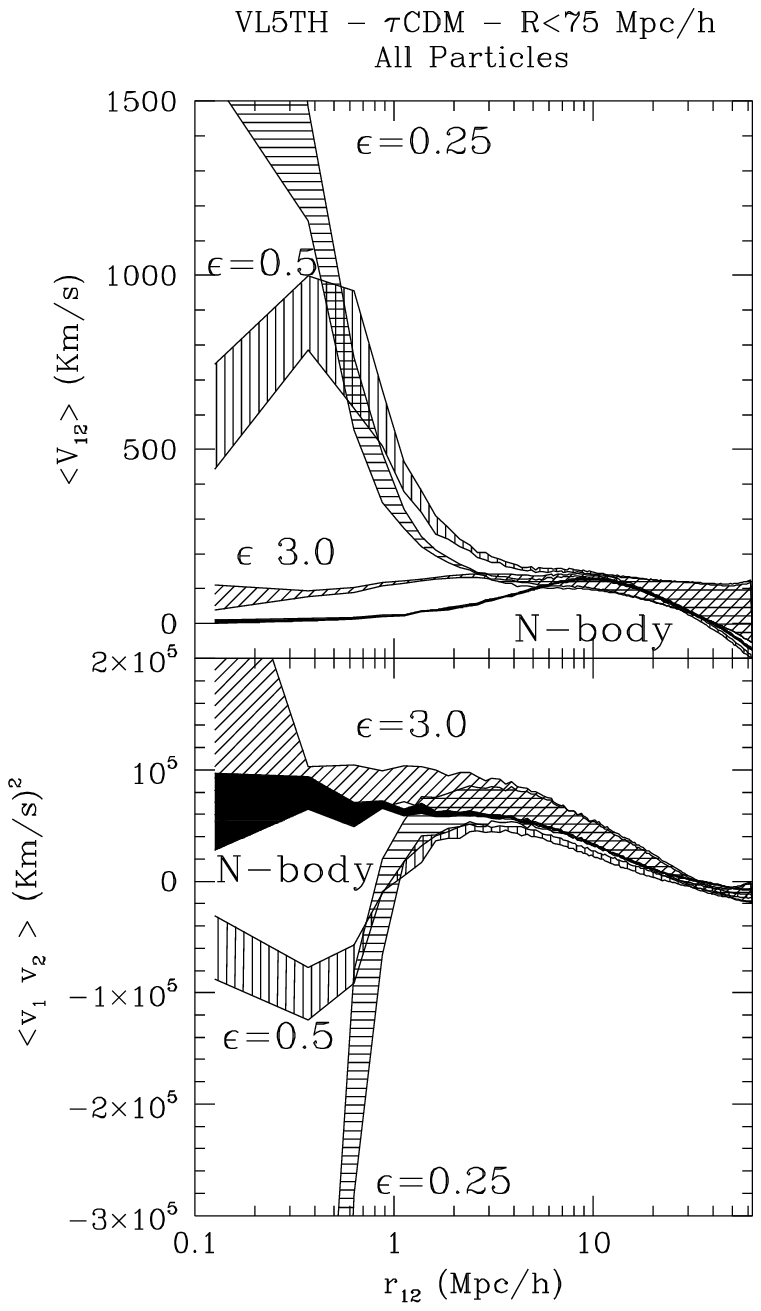}}
\caption{The same functions $\langle  V_{12} \rangle$ (upper panel) and 
$\langle  V_1 V_2 \rangle$  (lower panel)
displayed in fig.~\ref{fig:vcorr_FAMz_v5TH}  are plotted here for the FAMz case
and for three different values of the softening parameter.
Horizontally, vertically and diagonally dashed strips refer to $\epsilon=0.25, \ 0.5, \ 3.0 \ \hmpc$,
respectively.  
Note that the scales of the Y-axes are different from  fig.~\ref{fig:vcorr_FAMz_v5TH}.
}
\label{fig:vcorr_FAMz_v5THe}
\end{figure}

As shown in fig.~\ref{fig:vcorr_FAMz_v5TH}, the analysis of the
average relative pairwise velocity  (upper panel) and the 
{\it v-v}
correlation function
(lower panel) demonstrates that the main
differences between FAM and $N$-body velocity fields are on scales
smaller than $\sim 5 \hmpc$.  Indeed, at small relative  separations $\langle V_{12}
\rangle_s > \langle V_{12} \rangle_x$ and both are larger than the
$N$-body ones, showing that FAM systematically over-predicts the
relative infall velocity.  This spurious infall pattern appears
because of the small softening $\epsilon=0.5\hmpc$.
  As shown in
fig.~\ref{fig:vcorr_FAMz_v5THe}, varying the softening parameter has a
large effect on the correlation properties of the FAM velocity field.
With a smaller  softening of  $\epsilon=0.25 \hmpc$ (horizontally
dashed strips) the infall pattern becomes more prominent on small
scales.  Increasing the softening to $\epsilon=3.0 \hmpc$ (diagonally
dashed strips) reduces spurious infall, and
yields correlation properties which are very close to the $N$-body
ones over all scales.

\subsection{FAMz vs. Linear Theory and PIZA: Volume-Limited case}
\label{sec:flpr}

In this Section we compare the FAMz velocities
obtained in the VL experiments of Section~\ref{sec:volu}
to those predicted by linear theory and Zel'dovich approximation
using the same VL-catalogs.
Linear theory predictions have been obtained  from the 
gravity force computed from the particle distribution in s-space at the final time,
$\vg_{0,i}^s$: $\vv_i=H_{0}f_{0}\vg_{0,i}^s$.
Note that no attempt is made here to account for s-space distortion effects
which could be minimized using iterative procedures (e.g. Yahil \etal 1991).
Velocities in the Zel'dovich approximation have been obtained
by running FAMz with $n_{max}=1$ basis function which corresponds to 
straight line orbits
$\vx_{i}(D)=H_0^{-1}\vs_{{_{i,0}}} + D \vC_{i,1} -f_0 \cpar_{i,1} {\hat \vs}_{_{i,0}}$.
These velocities should be similar to those that would be obtained by applying
the PIZA method of Croft \& Gazta\~{n}aga  (1998) in  s-space.

\begin{figure}
\vspace{13truecm}
{\includegraphics{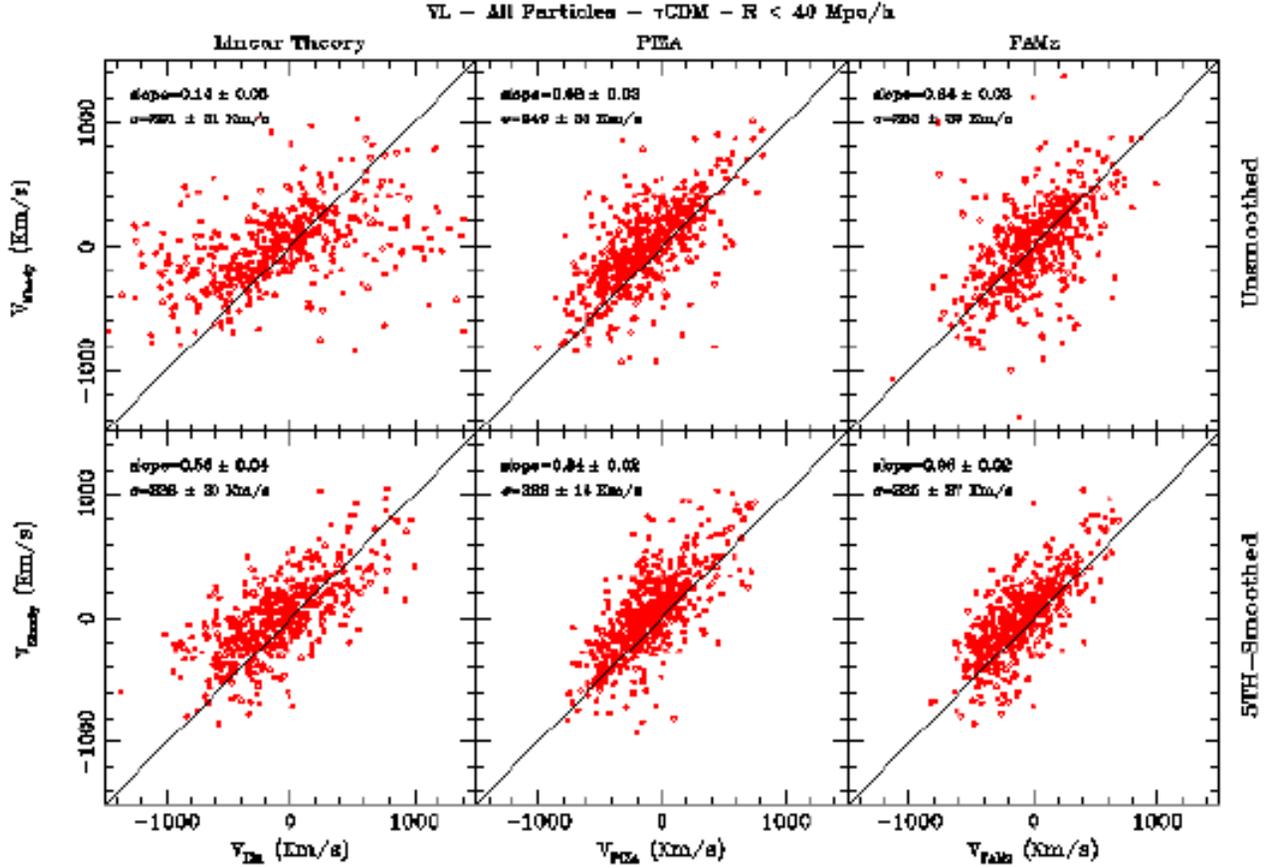}}
\caption{$N$-body velocities vs. peculiar velocities predicted by 
linear theory (left),
PIZA (middle) and FAMz (right) for $\sim 500$ objects randomly selected 
from the 50,000 particles in the VL-$\tau$CDM  catalogs.
In the upper plots the comparison is performed using unsmoothed peculiar 
velocities. The predicted velocities in the lower panels have been
smoothed  with a 5TH filter and are compared with unsmoothed $N$-body velocities.
The results of the fits are shown in all the plots.
The straight lines at $45^{\circ}$  are drawn to guide the eye.}
\label{fig:vv_FAMz_l_p_v}
\end{figure}

Fig.~\ref{fig:vv_FAMz_l_p_v} is similar to fig.~\ref{fig:vv_FAMz_v}.
The scatter plots in the upper panels
show the comparisons between $N$-body velocities
and velocities predicted using linear theory (left), PIZA (middle)
and FAMz (right) for 500 objects randomly selected from the 50,000 
particles in the 5  VL-$\tau$CDM catalogs. 

The scatter plot on the top left panel
appears to be dominated by a large number of out-liers
which cause linear theory predictions to 
overestimate peculiar velocities.
This is not surprising and derives from sparse sampling,
breakdown of linear theory in high density environments
and from the fact that s-space distortions were not accounted for.
To quantify the mismatch between $N$-body and linear velocities       
we have performed a linear regression of $N$-body velocities on model predictions
using all particles in each of the 5 VL-$\tau$CDM catalogs and assuming errors 
on the Y-axis only. 
The average slope, 1-D scatter of the regression and their rms values are 
shown in the panel and listed in table~\ref{tab:comp_tcdm} along with the
results of the other velocity models.
The slope of the best fitting line is very shallow and 
indicates that linear theory indeed severely overestimates peculiar velocities.
The 1-D velocity scatter, however, is rather small compared to the 1-D $N$-body velocity 
dispersion ($\approx 350 \kms$).
The situation improves considerably when using either the Zel'dovich approximation
(top central) or FAMz (top right). In both cases the slope of the best fitting line 
increases dramatically, especially for FAMz. The scatter 
around the fit, instead, remains almost unchanged.

Better results can be obtained by smoothing model velocities on small 
scales, where all the approximations break down. This is demonstrated by the
scatter plots shown in the bottom panels of  fig.~\ref{fig:vv_FAMz_l_p_v}
in which the unsmoothed $N$-body velocities are compared with
peculiar velocities predicted by  linear theory
(bottom left), PIZA (bottom central) and FAMz (bottom right) 
all of them smoothed with a Top Hat filter of radius $5 \hmpc$.
In all cases the match between true and predicted velocities improves considerably,
especially for the case of linear theory.
However, only FAMz returns unbiased velocity predictions.

Two sources of errors mainly affect velocity reconstruction 
methods: sparse sampling and the breakdown of model predictions in the 
highly nonlinear regime.
Both uncertainties affect all velocity-velocity 
comparisons of fig.~\ref{fig:vv_FAMz_l_p_v}
(and those in fig. 3 of Nusser \& Branchini (2000)), 
and therefore do not affect the relative comparison of the 
three reconstruction methods. 
However, these errors need to be accounted 
for to estimate absolute errors or 
when comparing our results with those of other tests that
adopt similar velocity models but use different mock catalogs, 
as in  Croft \& Gazta\~{n}aga (1998) and Berlind, Narayanan \& 
Weinberg (2000).
For example, 
the average inter-particle separation of 4.75 $\hmpc$
in our mock VL catalogs is similar to the galaxy-galaxy 
separation in the inner part of redshift catalogs like PSC$z$, 
2dF (Colless \etal 2001) and SDSS (Friemann \& Szalay 2000)
but is larger than that of particles in 
mock catalogs used to test most velocity models. 
Furthermore, as explained in  Section~\ref{sec:mock},
our mock catalogs contain spurious pairs and 
triplets of nearby objects. Their presence 
artificially increases the amplitude of model velocities
in high density environments and amplify the magnitude of 
the reconstruction errors.
Both effects conspire in making our error 
estimates larger than those obtained from more ideal mock catalogs,
like those used in the analyses of
of Croft \& Gazta\~{n}aga (1998) and Berlind, Narayanan \& 
Weinberg (2000).

To check the dependence of our results on the
cosmological model we have repeated the same analysis 
using the 5 VL-$\Lambda$CDM catalogs.
The results of the fits to the {\it v-v}
comparisons are listed in table~\ref{tab:comp_tcdm}.
The performances of PIZA and FAMz in a $\Lambda$CDM universe
are similar to those obtained in  a $\tau$CDM one,
indicating that the goodness of the velocity reconstruction
depends little on the underlying cosmological model.
The failure of linear theory, however, is more severe  
in the $\Lambda$CDM universe. This is a consequence of
the fact that this cosmological model is characterized by
a  larger number of high density peaks in which linear theory
breaks down.

Smoothing FAMz velocities with a 5TH filter 
returns unbiased 
velocity predictions in both $\tau$CDM and 
$\Lambda$CDM scenarios.
Increasing the smoothing length would have the effect 
of bring PIZA velocities (and, eventually, the linear ones)  
into agreement with the $N$-body ones.
However, as pointed out by,
Berlind, Narayanan \& Weinberg (2000), 
increasing the smoothing scale would also cause the FAMz velocity to be underestimated,
as a consequence of the fact that errors in predicted velocities
correlates with the predicted velocities themselves.
We conclude that $5 \hmpc$ should be regarded as the optimal smoothing scale for FAMz
velocities.

\begin{table}
\centering
\caption[]{$N$-body velocities vs. linear theory, PIZA and FAMz velocity predictions
for the VL (upper part) 
and FL (lower part) experiments and for two different cosmological models
($\tau$CDM and $\Lambda$CDM).
The average slope, $b$, and 1-D velocity scatter  $\sigma$ (in $\kms$) are shown along with 
their rms scatter, measured over all particles in 
the 5 independent experiments.
The best fit values are listed for the case of 
unsmoothed predicted velocities (left hand side) and 5TH-smoothed predicted velocities
(right hand side).} 
\tabcolsep 2pt
\begin{tabular}{cccccccc} \\  \hline
 & & & Unsmoothed  &  &  & 5TH-smoothed &   \\ \hline 
Model &   & Linear   & PIZA & FAMz  & Linear & PIZA & FAMz \\ \hline
VL-$\tau$CDM & $b$ & $0.14\pm 0.05$ & $0.58 \pm 0.03$  & $0.64 \pm 0.03$  & $0.56 \pm 0.04$ & $0.84\pm 0.02$ & $0.96 \pm 0.02$ \\
& $\sigma$& $291 \pm 31$ & $240 \pm 30$ & $253 \pm 33$ &  $238 \pm 30$ & $222\pm 14$ & $225 \pm 27$  \\ \hline
VL-$\Lambda$CDM & $b$ & $0.03\pm 0.02$ & $0.51 \pm 0.07$ & $0.58 \pm 0.04$ & $0.26 \pm 0.04$ & $0.81\pm 0.05$ & $0.98 \pm 0.06$  \\
& $\sigma$ & $320\pm 50$ & $285 \pm 40$  & $295 \pm 60$ & $280 \pm 57$ & $ 254 \pm 35$ & $271 \pm 56$  \\ \hline \hline
FL-$\tau$CDM & $b$ & $0.13\pm 0.02$ & $0.52 \pm 0.07$  & $0.74 \pm 0.04$ & $0.65 \pm 0.05$ & $0.93\pm 0.04$ & $1.03 \pm 0.04$ \\
 & $\sigma$& $291\pm 23$ & $249 \pm 25$ & $242 \pm 16$ & $232 \pm 18$ & $223\pm 11$ & $213 \pm 9$    \\ \hline
FL-$\Lambda$CDM & $b$ & $0.04\pm 0.01$ & $0.47 \pm 0.06$  & $0.57 \pm 0.06$ & $0.28 \pm 0.06$ & $0.88\pm 0.04$ & $1.01 \pm 0.04$ \\  
 & $\sigma$& $270\pm 60$ & $210 \pm 26$ & $232 \pm 31$ & $223 \pm 52$ & $ 192 \pm 28$ & $209 \pm 23$ \\ \hline
\end{tabular}
\label{tab:comp_tcdm}
\end{table}

\section{Tests with Flux-Limited Catalogs}
\label{sec:Nbodyflux}

Real datasets 
typically consists of flux (or magnitude) limited catalogs.  
Here we investigate how well
FAM performs when applied to the more realistic mock flux limited
catalogs described in Section~\ref{sec:mock}.  We will be concerned
with flux limited samples and will neglect other possible sources of
systematic errors like  unobserved regions of the sky (e.g.  Zone of
Avoidance), the occurrence of morphology and  density biases and so
on.

In all experiments we have chosen to use the the selection function of
PSC$z$ galaxies (eq.~\ref{eq:self}) since that catalog still
represents the deepest all-sky redshift survey presently available.  The mass of
each mock galaxy is equal to the inverse of the PSC$z$ selection
function measured at the galaxy's x-space location.  This may generate
systematic errors when the FAM reconstruction is performed s-space
(i.e. with FAMz) which are included in the error budget 
in the following two Sections.

\subsection{Unsmoothed Flux-Limited: FL}
\label{sec:Nbodyfluxu}

We have applied FAMz to the unsmoothed, flux limited catalogs
labeled FL in table~\ref{tab:mock} and 
displayed in the upper-right panel of fig.~\ref{fig:maps}. 
In this Section we present the results of  {\it v-v} analyses 
analogous to those in 
Section~\ref{sec:volu}.

\begin{figure}
\vspace{13truecm}
{\includegraphics{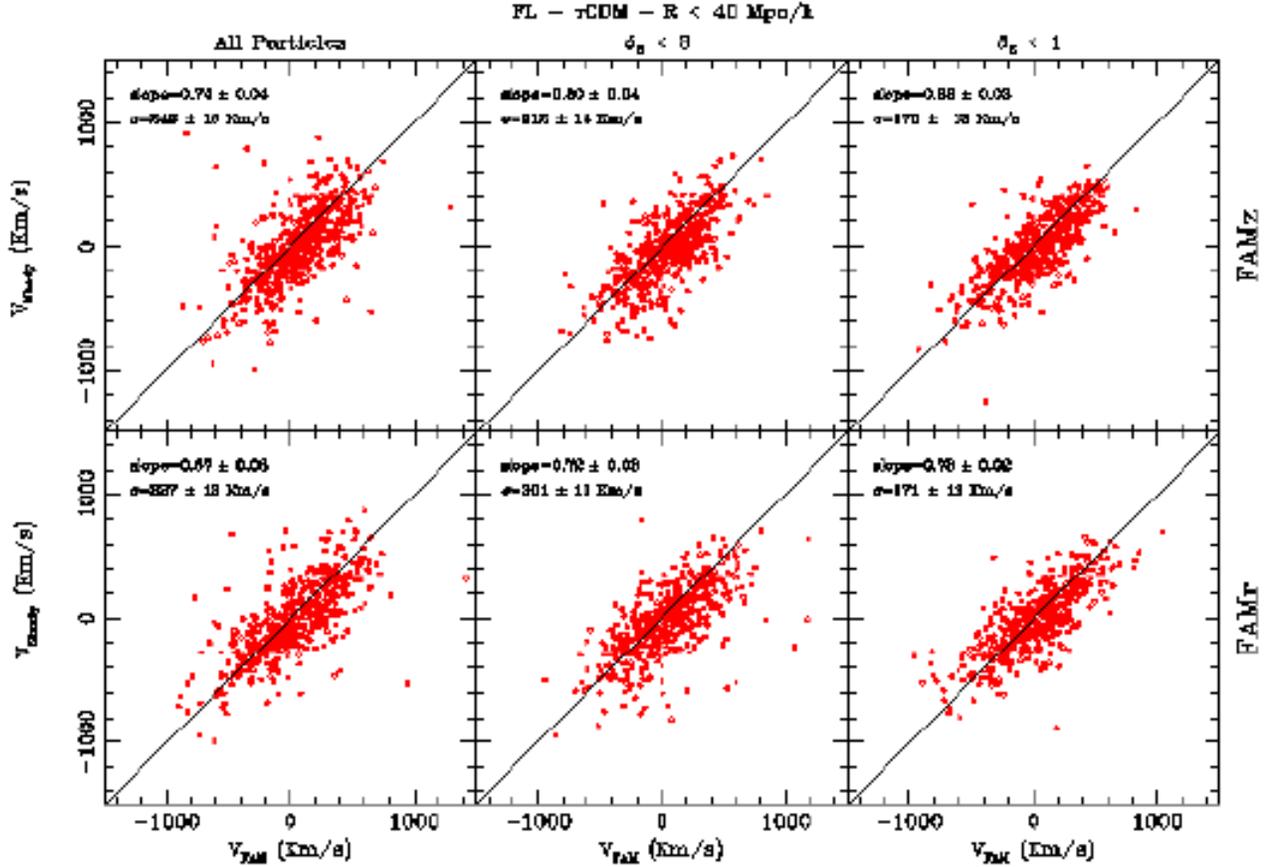}}
\caption{$N$-body vs. FAM velocity scatter plots in x- (bottom
panel) and s-space (top panel) for 500 randomly selected objects
from the 5 FL-$\tau$CDM  catalogs.  Panels on the left: all particles.
Panels in the middle: particles with $\delta_5 < 3$.  Panels on the
right: particles with $\delta_5 < 1$.  The straight line shows the
expected correlation.  The results of the fit are shown in all the
panels.  }
\label{fig:vv_FAMz_flim}
\end{figure}

The {\it v-v} scatter plots for  FL-$\tau$CDM catalogs are
shown in fig.~\ref{fig:vv_FAMz_flim}, both for FAMz (upper panels) and
FAMr (lower panels), for three different overdensity cuts.  The
scatter plots look similar to those of the volume limited tests
(fig.~\ref{fig:vv_FAMz_v}). 
The parameters of the linear regression of $N$-body velocities on FAM 
which are listed in the lower part of 
table~\ref{tab:tcdm} confirm the visual impression 
that FAM velocities are systematically
larger than the $N$-body ones.
Moreover, FAMr produces systematically larger velocities
than FAMz. The mismatch and between FAM and $N$-body velocities
decreases outside high density environments.
For $\delta_5 < 1$, FAMr and FAMz velocities 
are only $10 - 20 \%$ larger than the $N$-body ones and the mismatch 
further decreases when performing the $4\sigma$ clipping procedure described in 
Section~\ref{sec:volu}.
The analogy with the results found in is Section~\ref{sec:volu}
shows that selection effects do not seem
to affect FAM performances too much.  The most
striking feature is perhaps that the scatter around the fit
is very similar to that found in the volume limited
experiment of Section~\ref{sec:volu}. 
FAM performances are similar in the $\tau$CDM and $\Lambda$CDM cosmological models
despite of their very different 
values of $\sigma_8$, and  greatly improve when
high density environments
are excluded from the analysis. 

\begin{figure}
\vspace{13truecm}
{\includegraphics{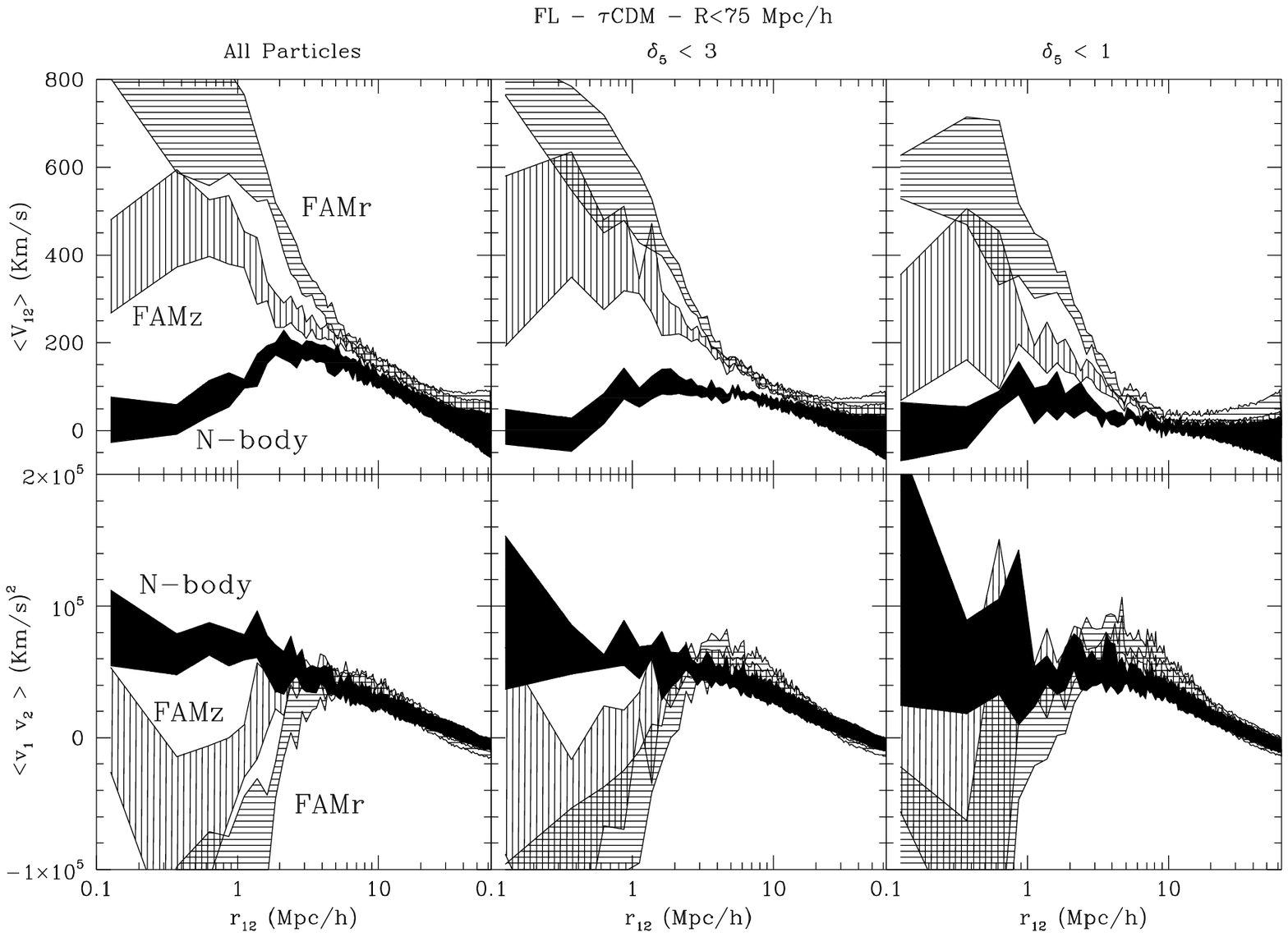}}
\caption{$ \langle V_{12} \rangle$ (upper panels), and 
$ \langle V_1 V_2 \rangle$  (lower
panels), as a function of the pair separation $r_{12}$ from 5 FL-$\tau$CDM 
experiments.  Vertically-dashed strip: FAMz.
Horizontally-dashed strip: FAMr.  Dark strip: $N$-body.  Left panels:
all the particles are considered.  Central panels: particles with
$\delta_5 < 3 $.  Right panels: particles with $\delta_5 < 1 $.  }
\label{fig:vcorr_FAMz_flim}
\end{figure}

Also the results of the velocity correlation analysis are quite
similar to those of the unsmoothed-volume limited tests, as it is
evident from the similarities between figs.~\ref{fig:vcorr_FAMz_v}
and \ref{fig:vcorr_FAMz_flim}.  However, selection effects are
responsible for three main differences.  First, the amplitude
of the difference  between FAM and $N$-body velocity correlations 
below $\sim 1 \hmpc$ is twice larger than in the volume-limited
case.  Second, the agreement with the $N$-body results shifts to a
somewhat larger scale. Third, the scatter around the mean
correlation value increases.

The results of this Section show that, when applied to mock PSC$z$
flux-limited catalogs, FAMz  produces almost unbiased estimates
of peculiar velocities outside high density regions. Moreover, the
correlation properties of the FAM and $N$-body velocity fields 
agree on scales larger than $ 5-8 \hmpc$. These results are remarkably
close to those of the volume-limited tests. 
 
\subsection{Smoothed Flux-Limited: FL5TH}
\label{sec:Nbodyfluxs}

In the previous sub-section  we have applied FAM to the 
particle distribution in the FL catalogs. 
Here we perform an ideal experiment analogous to that of 
Section~\ref{sec:vols} and show
the results of FAM applied to the FL5TH catalogs in which
the objects  sample a smoothed version of the 
particle distribution in the FL catalogs. 
A typical  FL5TH mock catalog is displayed in the 
bottom-right panel of fig.~\ref{fig:maps}. 
As for the case of the VL5TH
tests discussed in Section.~\ref{sec:vols}, the results
do not depend on the local density and therefore we only show the results 
relative to the full sample.

\begin{figure}
\vspace{13truecm}
{\includegraphics{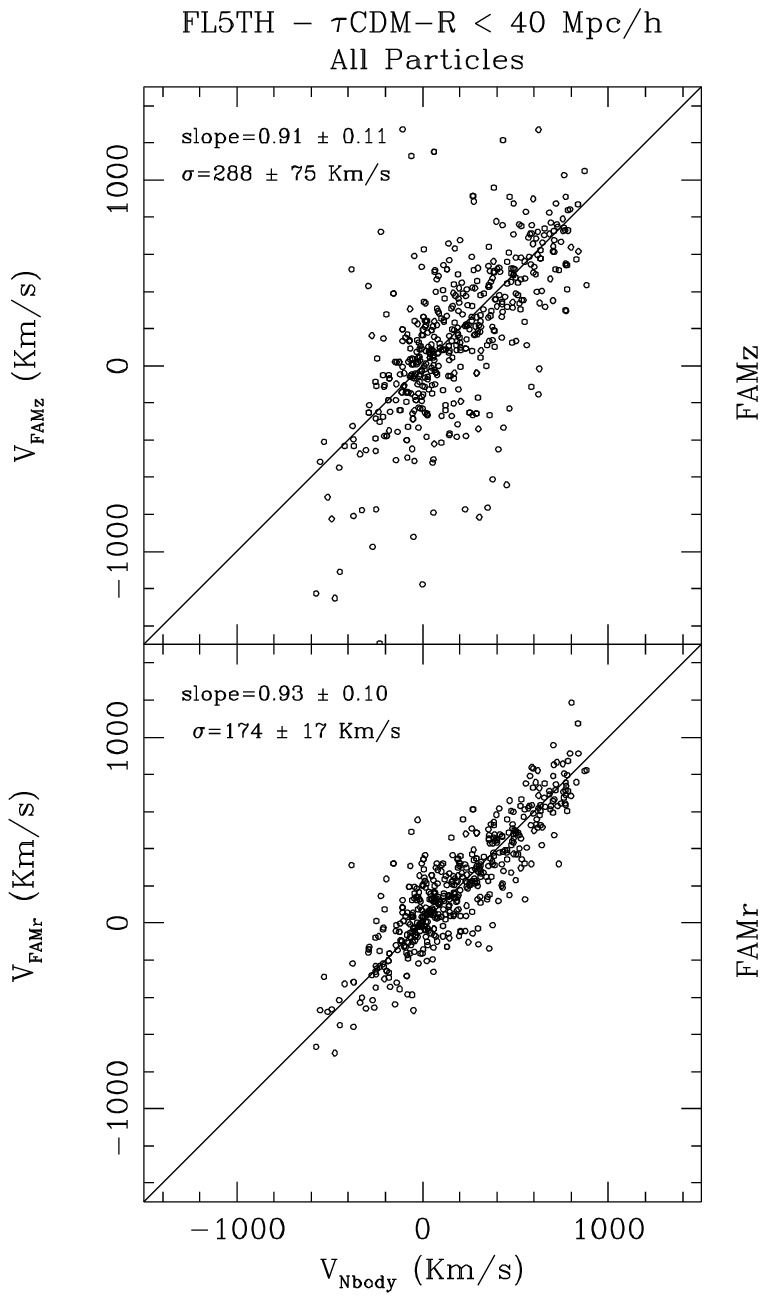}}
\caption{Same as fig.~\ref{fig:vv_FAMz_v5TH} but referring to the  
5 FL5TH-$\tau$CDM mock catalogs.  
}
\label{fig:vv_FAMz_flimx}
\end{figure}

Fig.~\ref{fig:vv_FAMz_flimx} shows the  {\it v-v} comparison for the
5 FL5TH-$\tau$CDM experiments.
As in Section.~\ref{sec:vols} we have performed 
linear regressions of FAM velocities on
$N$-body ones, where errors are associated to  FAM velocities only.
The results are
displayed in fig.~\ref{fig:vv_FAMz_flimx} and in
table~\ref{tab:tcdm_flimx}. They show that FAM peculiar
velocities are systematically smaller than the $N$-body ones.
Such discrepancy, however, is below the $1\sigma$ significance level.
The random errors are  larger than in all previous tests,
especially for the FAMz method, probably indicating that the
smoothing-resampling procedure enhances the amplitude of the shot
noise errors.  It is important to stress that, like in the
VL5TH tests, the results do not change when
applying different density cuts, which indicates that 
by applying the smoothing-resampling strategy 
there is no need of identifying
high density environments and exclude them from the {\it v-v}
analyses.  FAMr results seem to depend little on the underlying
cosmological model, as indicated by the results listed in the lower
part of table~\ref{tab:tcdm_flimx}.  
This is not true for FAMz,
indicating that  in a universe with a
large value of $\sigma_8$, characterized by the presence of many high
density peaks, some of them enhanced by effect of the shot noise
and smeared by s-space distortions, the
smoothing-resampling procedure might not be sufficient in providing
FAMz-velocity predictions which are uniform throughout the volume of
the sample.

\begin{table}
\centering
\caption[]{FAMr and FAMz results for FL5TH experiments 
in the $\tau$CDM (upper part) and $\Lambda$CDM (lower part) cosmological models.
The average slope, $b$, and 1-D velocity scatter,  $\sigma$ (in $\kms$),
measured over the 5 independent experiments
are shown along with  their rms scatter.} 
\tabcolsep 2pt
\begin{tabular}{ccccc} \\  \hline
             &         & FAMr   &  FAMz                          &   \\ \hline 
Model        &         &   All            &   All           &  \\ \hline
FL5TH-$\tau$CDM    & $b$     & $0.93\pm 0.10$   &  $0.91\pm 0.11$ &  \\
             & $\sigma$& $174\pm 17$      &  $288\pm 75$    &  \\ \hline
FL5TH-$\Lambda$CDM & $b$     & $0.93\pm 0.07$   & $0.99\pm 0.20$  &  \\
             & $\sigma$& $142\pm 27$      &  $ 256 \pm 134$ &  \\ \hline
\end{tabular}
\label{tab:tcdm_flimx}
\end{table}

\begin{figure}
\vspace{13truecm}
{\includegraphics{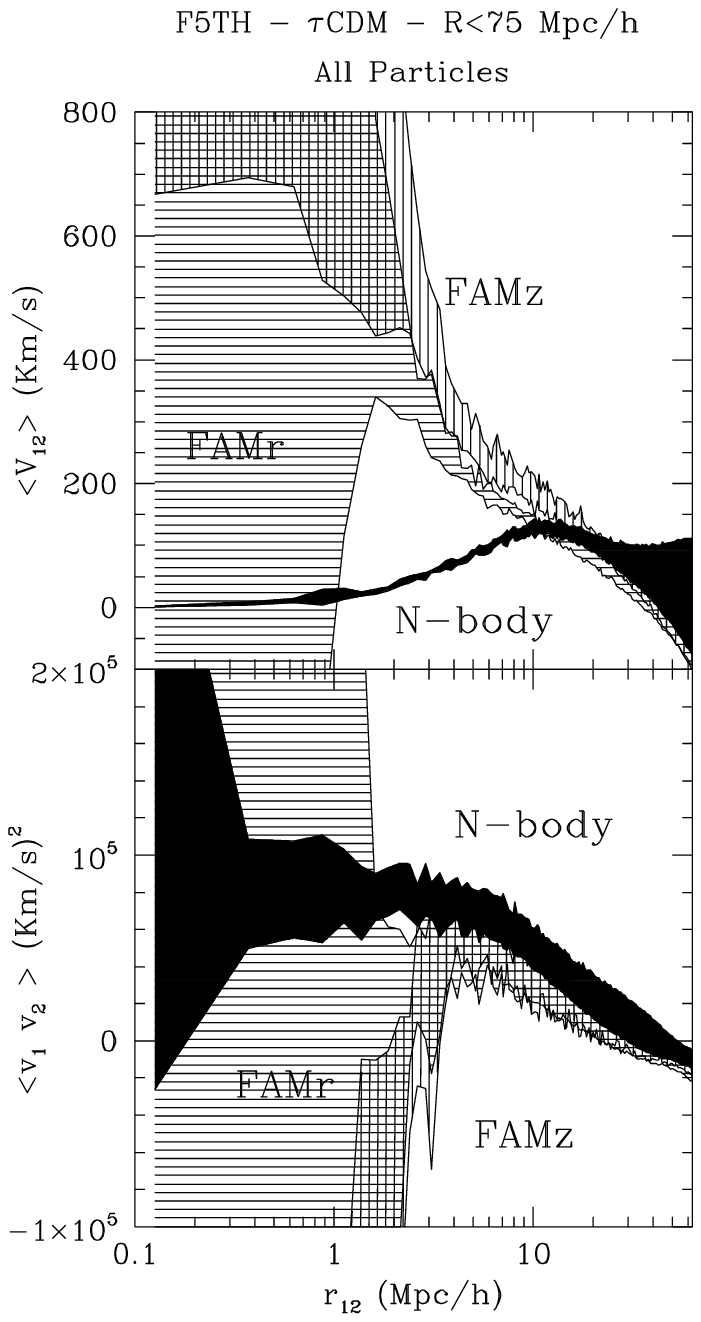}}
\caption{Same as fig.~\ref{fig:vcorr_FAMz_v5TH} for the
case of  5 FL5TH-$\tau$CDM experiments.
}
\label{fig:vcorr_FAMz_flimx}
\end{figure}

The effect of shot-noise is particularly dramatic for FAM velocity
correlations.  As seen in
fig.~\ref{fig:vcorr_FAMz_flimx}, the random uncertainties in the
predicted velocity correlations become extremely large below the
scale of $\sim 2 \hmpc$.  On scales larger than $2 \hmpc$ the results  
look similar to those obtained from the flux limited tests, 
displayed in the plots on the left hand side of
fig.~\ref{fig:vv_FAMz_flim}. In particular, the 
$1\sigma$ uncertainty strip of the FAM-predicted $\langle V_{12} \rangle$ 
statistics overlaps with the $N$-body result on scales larger than 
 $\sim 8 \hmpc$. A similar behavior also characterizes 
the $\langle V_1 V_2 \rangle $ statistics. In this case, 
however, the agreement between FAM and $N$-body is already found down to scales 
as small as  $\sim 3 \hmpc$.

\subsection{FAMz vs. Linear Theory and PIZA: Flux-Limited case}
\label{sec:flpf}

In this section we repeat the analysis performed in Section~\ref{sec:flpr}
and compare the FAMz velocities discussed in 
Section~\ref{sec:Nbodyfluxu} with those predicted by 
linear theory and PIZA in s-space, using the same FL catalogs.

\begin{figure}
\vspace{13truecm}
{\includegraphics{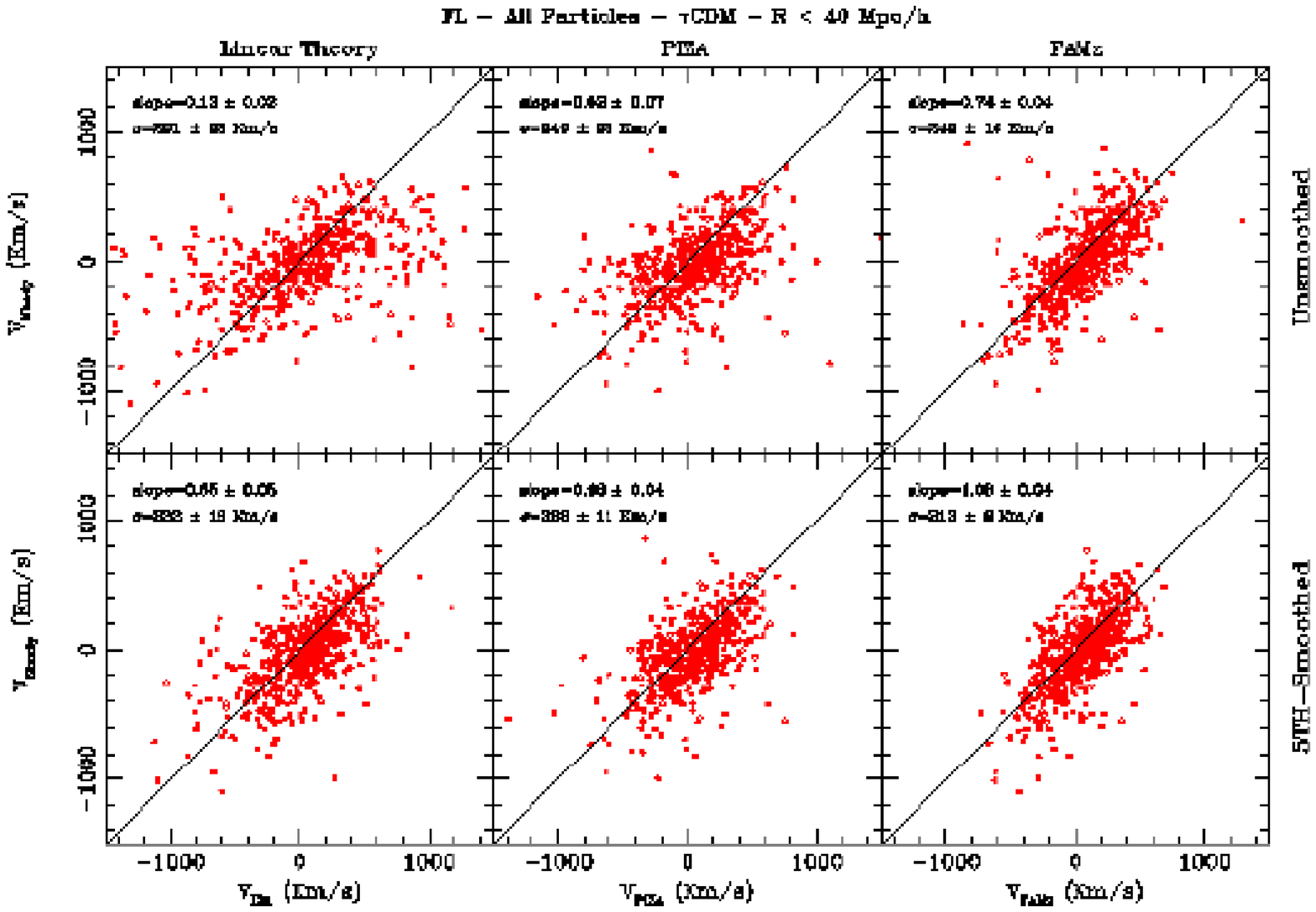}}
\caption{Same as fig.~\ref{fig:vv_FAMz_l_p_v} for the case of  FL-$\tau$CDM  catalogs.}
\label{fig:vv_FAMz_l_p_f}
\end{figure}

The results for the 5 FL-$\tau$CDM catalogs are displayed in 
fig.~\ref{fig:vv_FAMz_l_p_f} which is analogous 
to fig.~\ref{fig:vv_FAMz_l_p_v} and show the 
{\it v-v} comparison between unsmoothed $N$-body velocities
and velocities predicted by
linear theory (top left), PIZA (top central) and FAMz (top right). 
Predicted velocities in the lower panels were smoothed
using a Top Hat filter with an adaptive radius $R_{TH}={\rm Max}[5,l] \hmpc$,
where $l$ is the average particle-particle separation at the given redshift.
The results of the linear regressions of $N$-body velocities on the
predicted ones are displayed in fig.~\ref{fig:vv_FAMz_l_p_f} and 
in the lower part of  table~\ref{tab:comp_tcdm}.

When no smoothing is applied to predicted velocities (upper panels),
all methods overestimate the peculiar velocities. The situation is thus similar
to that of the volume limited experiment performed in Section~\ref{sec:flpr}.
This bias is particularly severe in the case of linear theory and 
decreases significantly by applying a top hat
 smoothing of  radius $ R_{TH} \simeq  5 \hmpc$
to predicted velocities, as shown by the scatter plots in the 
bottom panels. In particular, FAMz smoothed velocity predictions are bias-free and 
with a random error of of $220 \kms$,
nearly independent on the underlying cosmological model. 
These results demonstrate that, unlike the two other method considered here, 
applying FAMz to a PSC$z$-like  redshift 
catalog allow to predict a model for the cosmic velocity field which is 
unbiased, provided that FAMz velocities are smoothed on a scale of 
$\sim 5 \hmpc$.

\section{Summary and Conclusions}
\label{sec:disconc}

We presented extensive tests of the FAM method for reconstructing
peculiar velocities from particle distribution in x- and s-space.
Using the analytic spherical collapse model and a suite of realistic
volume-limited and flux-limited mock catalogs, we showed that FAM can
successfully recover peculiar velocities from the particle
distribution in x- and s-space. 
Errors in the predicted velocities are mainly due to 
shot noise and highly nonlinear motions that FAM fails to 
model correctly. Shot noise errors depend on the sparseness of 
galaxy sampling and thus on the particular dataset considered. 
To obtain a realistic error estimate we set the number density of galaxies in 
our mock catalogs similar to that in the 
inner part of the real PSC$z$, 2dF and SDSS catalogs.
Errors due to nonlinear motions
can be reduced by avoiding high density regions.
Indeed, discarding objects in the highest density
peaks brings systematic errors down to a level smaller than $20\%$
and reduces the amplitude of random
uncertainties from $\sim 250 \kms$ to $\sim 150 \kms$ in the 
volume limited experiments.  
When applied to flux-limited mock catalogs mimicking the distribution of galaxies
in the PSC$z$ redshift survey, FAM systematic errors 
outside dense regions are still below 20 \% and the  
1-D random velocity errors amount to $120-170 \kms$.
Moreover,  FAM recovers
the correlation properties of the underlying velocity field down to
scales of $\sim 5 \hmpc$ in both  flux and volume limited cases.
In practice, it may be
difficult to measure the local density from sparse datasets, in order
to discard high density peaks. 
However, particles in these regions
have very large velocity residuals, and appear as out-liers in the
{\it v-v} scatter plots. Eliminating all 
points deviating by more than $4\sigma$ from the best fitting
line improves the results of the fit and returns velocity estimates which
are almost unbiased. 
Another strategy to reduce the impact of nonlinear motions is that of 
smoothing the model velocity field. We have found that FAMz velocities 
smoothed with a Top Hat filter of radius $ 5 \hmpc$ are free of systematic biases.
Clearly, there is nothing fundamental in this smoothing scale. To the contrary, the amount of
smoothing required to minimize systematic errors depends on
the strength of nonlinear motions, sampling rate and on the particular
method used to compare model and observations.
Other reconstruction methods like those based on
linear theory and Zel'dovich approximation could also predict unbiased
velocities, provided that an appropriate smoothing filter is applied.
However, the smoothing scale required by FAMz to eliminate 
systematic errors is smaller than that required  by 
linear theory or PIZA and thus allow to retain more information
which lead to a more precise estimates of cosmological 
parameters such as $\beta=\Omega_m^{0.6}/b$.

Since FAM errors depend on 
several external factors, like the particular dataset
considered, it is difficult to 
compare our results with those of other analyses.
For example, it is somewhat surprising that 
errors in the Cartesian components of FAMz velocities  
predicted from our mock PSC$z$ catalogs
are a factor of two larger 
than those of the velocities
obtained by applying the
the iterative, quasi-linear model 
of Sigad \etal (1998)  to the IRAS 1.2 Jy. catalog
(see Willick \etal 1997).
Indeed, the better treatment of nonlinear dynamics 
by FAMz and the denser sampling of the PSC$z$ catalog
should reduce these uncertainties 
rather than increase them. 
As discussed by Branchini \etal (2001), this apparent inconsistency
derives from the differences in the mock catalogs used to calibrate the 
two reconstruction methods. 
Peculiar velocities in our mock catalogs, extracted
from the AP$^3$M simulations of Cole \etal (1998),
are more nonlinear than those of the mock catalogs 
used Willick \etal (1997) that were extracted
from PM $N$-body simulations.
The higher``temperature'' of the velocity field in our mocks
increases the uncertainties in the velocity reconstruction 
procedure. Errors in velocity predictions are further amplified by the
spurious pairs and triplets of nearby particles
contained in our mocks. The net result is that
our estimated FAMz velocity errors are larger that
those of Willick \etal (1997). 
Another example is given by the large systematic errors 
affecting our linear velocity predictions compared to FAM velocities,
as shown by figures~\ref{fig:vv_FAMz_l_p_f} and~\ref{fig:vv_FAMz_l_p_v}.
This result suggests that estimates of $\beta$
from $v-v$ comparisons that use a linear model velocity field based on the 
PSC$z$ catalog (e.g. Nusser \etal (2000) and 
Branchini \etal (2001)) are biased low
and that this bias could be reduced by using the
FAMz model, instead.
A quantitative assessment of this bias and its reduction, 
however, needs to
account for the differences between the various models and analyses.
First of all the linear model used by 
Nusser \etal (2000) and Branchini \etal (2001)
was corrected for redshift s-space distortions
while our model is not.
Second, in this work the $v-v$ comparisons are performed 
through simple linear regressions
while the VELMOD analysis of Branchini \etal (2001)
and the mode-by-mode comparison of Nusser \etal (2000)
are less trivial and require additional 
manipulation of the data that may affect the estimate of 
$\beta$. While we plan to quantify these effects in a future
paper we also point out that the tests performed by Branchini \etal (2001)
and Nusser \etal (2000) have indicated that possible systematic errors  
on $\beta$ are below the 20 \% level.
The two previous examples illustrate  
that the results presented in this work 
are only valid for volume limited and flux limited
samples that mimic the PSC$z$ catalog.
As for any  reconstruction method, random and systematic errors
need to be estimated again when a new dataset is considered
or a new model vs. data analysis is implemented.
Such error analysis is best done by applying FAM to a set of 
realistic mock catalogs, as we did in this work

The dependence of the clustering properties on the type of real
galaxies implies that most galaxies are biased tracers of the
underlying mass fluctuations (Loveday \etal 1995, Baker \etal 1998,
Hawkins \etal 2001, Norberg \etal 2001).  NB have shown that a local
biasing scheme can be easily incorporated in FAM by assigning to each
particle a mass proportional to ${\cal W}=(1+\delta)/(1+\delta^g)$
where $\delta$ and $\delta^g$ are, respectively, the mass and galaxy
density contrast in the vicinity of the particle.  Alternatively,
given $\cal W$ one can obtain the mass density field from the galaxy
distribution. The mass density field can then be sampled by a discrete
particle distribution to be used as input in FAM, instead of the
original biased particle distribution.  Any biasing relation is
naturally defined in real space; hence an application of FAM on a
distribution of galaxies in s-space must be done iteratively, if
biasing is to be incorporated properly.  Nevertheless, several authors
have shown that forcing a biasing relation in s-space does not
introduce large errors (e.g. Verde \etal 1998, Scoccimarro \etal 2001,
Szapudi 1998, Sigad, Branchini \& Dekel 2000, Narayanan \etal 2001).
By making reasonable assumptions on the statistical properties of the
mass density, it is even possible to derive the biasing relation in
redshift space directly from the redshift catalog. For example, a
biasing scheme can be obtained by mapping the probability distribution
function (PDF) obtained from the galaxy distribution into a mass PDF
which is assumed to be log-normal (Sigad, Branchini \& Dekel 2000) or
by extensively applying hybrid reconstruction techniques (e.g.
Narayanan \etal 2001).  In future work, we will test FAM using
different biasing schemes and also using mock galaxy catalogs
extracted from ``galaxies'' identified in $N$-body simulations using
semi-analytic models  (e.g., Kauffmann, Nusser \& Steinmetz
1997, Kauffmann \etal 1999, Benson \etal 2000).

Methods based on the least action principle, like FAM, can serve as a
time machine for recovering the particle positions and velocities at
any early epoch.  The particle distribution can then be filtered to
obtain the density field as a function of time. A backward
reconstruction of the large scale structure can be very rewarding. It can
provide an estimate for the initial density field and this could
discriminate among different cosmological scenarios, e.g. by detecting
non-Gaussian features in the primordial density field or by measuring
its power spectrum (Nusser, Dekel, \& Yahil 1995, Kolatt \etal 1996,
Monaco \etal 2000, Narayanan \etal 2001).  The performance of FAM as a
time machine has only been tested on the scale of the Local Group in
real space with no biasing (Peebles 1990, 1994), or with a simple halo biasing
prescription (Dunn \& Laflamme 1993, Branchini \& Carlberg 1994).
Therefore, there is a need for thorough testing of FAM to establish
its reliability at recovering the past orbits in redshift space,
over scales of cosmological interests and using realistic,
time-dependent biasing schemes (Nusser \& Davis 1994, Fry 1996,
Matarrese \etal 1997).  While we plan to address these issues in a
future paper it is important to note that our results, along with
those obtained by NB, show that FAMz performs better than 
linear theory and PIZA  (Croft \& Gazta\~{n}aga 1998) in
reconstructing peculiar velocities at the present epoch. It is
therefore reasonable to expect that the FAMz will also improve over
PIZA backward reconstruction and thus will be able to recover the
correct initial conditions down to scales smaller than $\sim 3 \hmpc$,
without enforcing any {\it prior} power spectrum like in the
Perturbative Least Action procedure (Goldberg \& Spergel 2000).

The results of our tests indicate that, once applied to the PSC$z$
catalog, FAMz will able to model the cosmic velocity field down to
scales of a few Megaparsecs with 1-D random uncertainty of $150 -200
\kms$, hence improving over presently available models also obtained
from the PSC$z$ catalog using either linear theory (Branchini
\etal 1999) or the PIZA method (Valentine, Saunders \&
Taylor 2000).  We plan to use this new velocity model to perform $v-v$
comparisons that should return an estimate of $\Omega_{m,0}$, with an
expected $1\sigma$ uncertainty of $10 - 15 \%$ and, because of the
wide range of scales probed by the recent data, to provide valuable
information on the scale dependence of the biasing relation.

In this paper,  FAM has been applied to systems containing up to
$N = 2\cdot10^4$ particles. Since the computational cost of FAM
reconstructions scales as $N \log N$, the method can also be applied to
new generation redshift surveys containing $10^5$ to $10^6$ objects,
like the 2dF and SDSS 
catalogs. In these new catalogs, however, the surveyed regions span
only a fraction of the celestial sphere.  This constitutes a potential
problem for reconstruction methods like FAM which require to be
applied to spherical catalogs to minimize the dynamical influence of
the neglected mass distributed outside the volume of the sample.  A
possible way to proceed is to reconstruct the velocity field over a
series of spherical sub-catalogs carved out of the parent survey.  The
sampling within these sub-catalogs will be denser than in the PSC$z$
catalog and the velocity model less affected by shot noise.  As a result,
applying the FAMz procedure to a set of spherical sub-samples filling
the volume of the parent catalog would return an unbiased model
velocity fields on unprecedented large scales and with 1-D random
uncertainties of about $100 \kms$. 
This would lead to a precise determination of the distribution of 
mass which could then be cross-correlated with absorption features 
in AGN spectra either due to the interviewing cold, neutral hydrogen 
(Ly$\alpha$ line) or to the warm-hot intergalactic medium (O$_{\rm VI}$
and  O$_{\rm VII}$ lines) to determine the ill-known baryon fraction at z=0. 
Another possibility would be to compare FAM velocity prediction 
with precise velocity measurements of galaxies in the very local universe, like
those  obtained from the Surface Brightness Fluctuation method 
(Tonry \etal 2000, 2001). This should allow to measure the value of $\Omega_{m,0}$
with an even higher precision.

\section{acknowledgments}
EB thanks the Technion of Haifa for its hospitality while part of this
work was done.   AN acknowledges
the support of a grant from the Norman and Helen Asher Space
Research Institute.
This research is supported by the Technion V.P.R Fund-Henri 
Gutwirth Promotion of Research Fund, and the German Israeli
Foundation for Scientific Research and Development.

\protect

\end{document}